\documentclass[acmtog]{acmart}

\AtBeginDocument{%
  \providecommand\BibTeX{{%
    \normalfont B\kern-0.5em{\scshape i\kern-0.25em b}\kern-0.8em\TeX}}}

\setcopyright{acmcopyright}\acmJournal{TOG}
\acmYear{2022}\acmVolume{41}\acmNumber{6}\acmArticle{1}\acmMonth{12} \acmDOI{10.1145/3550454.3555465}

\acmJournal{TOG}

\usepackage[boxed,ruled,linesnumbered]{algorithm2e} 

\SetCommentSty{mycommfont}
\let\oldnl\nl
\newcommand{\nonl}{\renewcommand{\nl}{\let\nl\oldnl}}

\SetAlFnt{\small}
\SetAlCapFnt{\small}
\SetAlCapNameFnt{\small}
\SetAlCapHSkip{0pt}
\IncMargin{-\parindent}

\newlength\savedwidth
\newcommand\whline[1]{\noalign{\global\savedwidth\arrayrulewidth
                               \global\arrayrulewidth #1} %
                      \hline
                      \noalign{\global\arrayrulewidth\savedwidth}}

\usepackage{booktabs} 
\usepackage{enumitem}
\usepackage{amsmath}
\usepackage{color, colortbl}
\definecolor{lightgray}{gray}{0.9}
\usepackage{xcolor}
\usepackage{wrapfig}
\usepackage[normalem]{ulem}
\usepackage[export]{adjustbox}
\usepackage[most]{tcolorbox}
\newtcolorbox{mybox}{colback=white,colframe=black,left=7 pt,top=5 pt,
bottom=0 pt,right=5 pt,boxsep=2 pt,toprule=0.5 pt,leftrule=0.5 pt, bottomrule=0.5 pt, rightrule=0.5 pt,breakable, enhanced, arc=0pt,outer arc=0pt}

\usepackage{romannum}
\usepackage{soul}

\begin{document}
\title{Computing Medial Axis Transform with Feature Preservation via Restricted Power Diagram}

\acmArticle{188}

\author{Ningna Wang}
\affiliation{%
  \institution{University of Texas at Dallas}
  \country{USA}
}
\email{ningna.wang@utdallas.edu}

\author{Bin Wang}
\affiliation{%
  \institution{Tsinghua University}
  \country{China}
}
\email{wangbins@tsinghua.edu.cn}

\author{Wenping Wang}
\affiliation{%
  \institution{Texas A\&M University}
  \country{USA}
}
\email{wenping@tamu.edu}

\author{Xiaohu Guo}\authornote{Corresponding author}
\affiliation{
  \institution{University of Texas at Dallas}
  \country{USA}
}
\email{xguo@utdallas.edu}

\renewcommand\shortauthors{Wang et. al}

\begin{abstract}
 We propose a novel framework for computing the medial axis transform of 3D shapes while preserving their \emph{medial features} via \emph{restricted power diagram} (RPD). Medial features, including \emph{external features} such as the sharp edges and corners of the input mesh surface and \emph{internal features} such as the seams and junctions of medial axis, are important shape descriptors both topologically and geometrically. However, existing medial axis approximation methods fail to capture and preserve them due to the fundamentally under-sampling in the vicinity of medial features, and the difficulty to build their correct connections. In this paper we use the RPD of medial spheres and its affiliated structures to help solve these challenges. The dual structure of RPD provides the connectivity of medial spheres. The surface \emph{restricted power cell} (RPC) of each medial sphere provides the tangential surface regions that these spheres have contact with. The connected components (CC) of surface RPC give us the classification of each sphere, to be on a medial sheet, a seam, or a junction. They allow us to detect insufficient sphere sampling around medial features and develop necessary conditions to preserve them. Using this RPD-based framework, we are able to construct high quality medial meshes with features preserved. Compared with existing sampling-based or voxel-based methods, our method is the first one that can preserve not only external features but also internal features of medial axes.
\end{abstract}

%
%



\begin{teaserfigure}
\center
  \includegraphics[width=\textwidth]{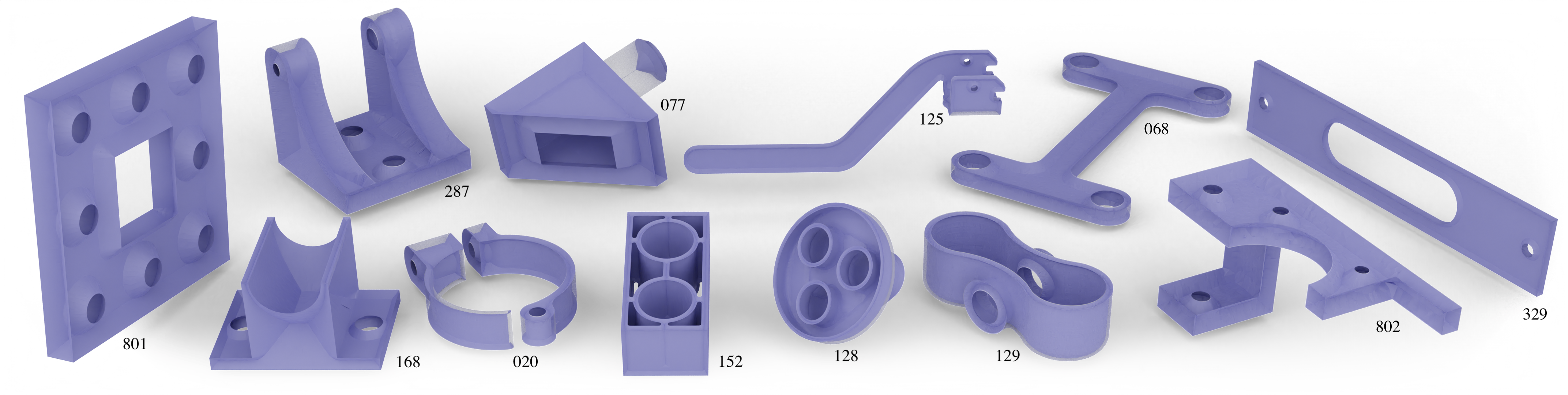}
  \caption{A gallery of our feature-preserving 3D medial axis results (in violet) from inputs of CAD meshes (in transparency) with sharp features.}
  \label{fig:teaser}
\end{teaserfigure}

\maketitle

\newcommand{\red}[1]{\textcolor{red}{#1}}
\newcommand{\blue}[1]{\textcolor{blue}{#1}}
\newcommand{\green}[1]{\textcolor{green}{#1}}
\newcommand{\brown}[1]{\textcolor{brown}{#1}}
\newcommand{\rspace}{\mathbb{R}}
\newcommand{\model}{\mathcal{S}}
\newcommand{\bmodel}{\partial \mathbf{S}}
\newcommand{\sample}{P}
\newcommand{\lfs}{LFS(\mathbf{x})}
\newcommand{\ma}{\mathcal{M}}
\newcommand{\mmesh}{\mathcal{M}_s}
\newcommand{\msphere}{\mathbf{m}}
\newcommand{\mcenter}{\boldsymbol \theta}
\newcommand{\anypoint}{\mathbf{x}}
\newcommand{\sharpangle}{\phi}
\newcommand{\powerdist}{d_{pow}}
\newcommand{\powercell}{\Omega^{pow}}
\newcommand{\vorcell}{\Omega^{vor}}
\newcommand{\powerseg}{\Gamma}
\newcommand{\rpc}{\mathbf{\omega}} 
\newcommand{\brpc}{\partial \mathbf{\omega}} 
\newcommand{\rps}{\gamma} 
\newcommand{\tanpl}{\mathbf{PL}}
\newcommand{\tanpoint}{\mathbf{p}}
\newcommand{\tannormal}{\mathbf{n}}
\newcommand{\ccl}{\mathbf{l}}
\newcommand{\cclpoint}{\mathbf{m}}
\newcommand{\cclcenter}{\mathbf{x}}
\newcommand{\cclnormal}{\mathbf{n}}
\newcommand{\ccldirection}{\mathbf{d}}
\newcommand{\energy}{E}
\newcommand{\numneighbors}{\mathcal{K}}
\newcommand{\ie}{\textit{i.e., }}
\newcommand{\eg}{\textit{e.g., }}
\newcommand{\pointP}{\mathbf{p}}
\newcommand{\powl}{\mathbf{l}_{pow}}
\newcommand{\tri}{\mathbf{t}}

\section{Introduction}

The medial axis \cite{blum1967transformation} is a fundamental geometric structure and has been widely used in approximating, simplifying, and analyzing shapes. The medial axis $\ma$ of a 3D shape $\model$ is simply defined as the set of centers of maximally-inscribed spheres touching two or more points on the surface $\bmodel$. Topologically, the medial axis $\ma$ is homotopy equivalent to $\model$. Geometrically, the medial axis  $\ma$ captures the protrusions and components of $\model$. The \textit{medial axis transform} (MAT) is the combination of the medial axis and the radius function defines on it. We represent MAT using a simplicial complex called \textit{medial mesh} $\mmesh$ and approximate the input 3D shape $\model$ by the union of enveloping volumes of the medial primitives (\emph{medial cones} and \emph{medial slabs}, see Sec.~\ref{sec:pre_ma}) of the medial mesh.

\begin{figure}[t]
    \centering
    \includegraphics[width=0.9\linewidth]{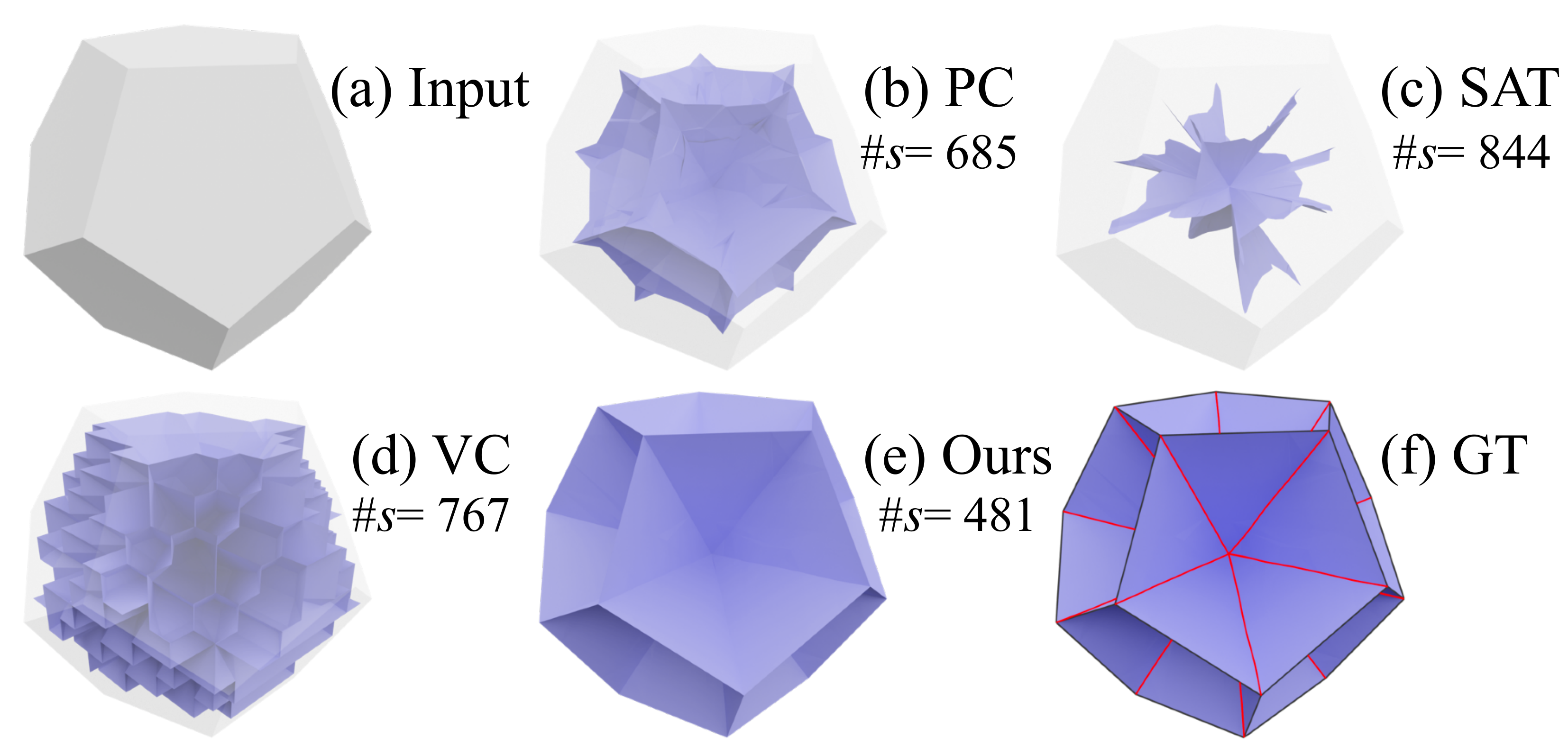}
    \vspace{-5pt}
    \caption{Approximations of medial axis of (a) a dodecahedron model computed with similar number of medial spheres (\#$s$), using (b) a sampling-based method Power Crust (PC) \cite{amenta2001power}, (c) another sampling-based method Scale Axis Transform (SAT) \cite{miklos2010discrete}, (d) a voxel-based method Voxel Core (VC) \cite{yan2018voxel}, and (e) our method. The ground truth medial axis (f) highlights external features in black and internal features in red.}
    \label{fig:intro_dode}
    \vspace{-5pt}
\end{figure}

We have observed that \textit{medial features} play a significant role in guaranteeing many topological and geometric properties of medial axis. \textit{External features} such as sharp edges and corners of the input 3D shape, which are common to CAD models, represent the non-smoothness of the surface (see Fig.~\ref{fig:intro_dode} (f) lines in black). It follows from the definition that the medial axis passes through the points where the surface is locally convex and non-smooth. \textit{Internal features}, on the other hand, defines the inner topological structure of seams and junctions (see Fig.~\ref{fig:intro_dode} (f) lines in red) so that medial axis has a natural decomposition into multiple manifold sheets. Applications such as hexahedral mesh generation of CAD models \cite{quadros2004laytracks} \cite{sampl2000semi} relies on internal features as a starting point to perform solid meshing. As a result, the medial features, both external and internal, constitute the foundation of medial axis as skeletal shape descriptors \cite{tagliasacchi20163d}.

However, all existing medial axis approximation methods that are designed to handle large input 3D meshes (\eg tens of thousands of triangle faces as shown in all models of Fig.~\ref{fig:teaser}) fail to preserve external features. For CAD models that commonly contain external features like convex sharp edges and corners, the approximated medial axis usually stops before touching these features, as shown in examples of Fig.~\ref{fig:intro_dode} (b)(c)(d), in which cases the structures of medial axis are not complete. For all methods depending on inner Voronoi balls of surface samples, no matter how dense the surface sampling points are, the density condition designed based on their \emph{local feature size} (LFS)~\cite{amenta2001power} can never be achieved at convex sharp edges and corners. This is because their LFS converges to zero near the external features, causing their sampling density to converge to infinity. Even though \emph{weak feature size} (WFS)~\cite{Chazal2005WFS} was proposed to bypass the issue of vanishing LFS for non-smooth shapes, there is no practical method yet to preserve those external features in the resulting medial axis. 

Furthermore, internal medial features are even more difficult to capture since they are hidden information of the input shape, as shown in Fig.~\ref{fig:intro_dode}. The essential reason of this failure is because of the under-sampling of medial spheres on internal features (red lines in Fig.~\ref{fig:intro_dode} (f)). Existing solutions tend to increase surface sampling rate (or decrease voxel size) globally in order to maintain the internal features. This eventually results in a large amount of non-feature medial spheres being sampled without any guarantee to capture the topologically-important internal features. 

In order to generate a medial mesh with feature preservation, we need to answer the following three questions: (1) How to sample and update feature spheres to guarantee their tangential properties? For example, spheres on medial seam are tangential to three points on the surface, while spheres on medial junction are tangential to four. (2) How to identify insufficient sampling on medial features, for both internal and external ones? (3) How to connect all medial spheres with correct topology and geometry? These open questions varies for these two types of medial features:

\begin{itemize}
    \item \textit{External features} can be trivially identified on the input mesh surface using the dihedral angle of two incident polygonal faces. We can sample zero-radius medial spheres on those convex external features. However, there are still three major challenges: (a) How to create the connectivity between these zero-radius feature spheres and their nearby non-feature spheres is still an open question. (b) The connectivity of neighboring feature spheres on external sharp edges could be broken by the ``invasion'' of nearby non-feature medial spheres. (c) Multiple convex and concave sharp edges could meet at a corner, where neighboring edges could form small angles, causing a ``corner cap'' that connects feature spheres at neighboring sharp edges. 

    \item \textit{Internal features}, on the other hand, are difficult to identify in the first place. It it known that with more surface samples used to construct inner Voronoi balls, the internal features are more likely to be well-captured. However, there is no indication to tell whether the sphere sampling on internal feature is sufficient or not, especially at the vicinity of external features in CAD models where the sampling is typically not dense enough.
\end{itemize}

In this paper, we present the first framework for computing an approximated MAT that preserves both external and internal medial features for an input 3D mesh surface while ensuring the approximation accuracy. Our method is based on a novel insight that the surface \textit{restricted power diagram} (RPD) can provide us auxiliary knowledge about the surface regions which medial spheres have tangential contacts with. This gives us the ability to classify medial spheres based on their contacts to the boundary surface (see details in Sec.~\ref{sec:classification}), and sample or update spheres to their ideal position by solving a quadratic energy optimization problem (as answers for question (1) above). We further develop the necessary conditions to preserve both external and internal features based on the information of \textit{restricted power cells} (RPCs) of these feature spheres with their neighbors (as answers for question (2) above). The \textit{restricted regular triangulation} (RRT, as dual of RPD) implies the connectivity \cite{amenta2001powerjournal} of discrete medial spheres to form a mesh approximation of the medial axis (as partial answers for question (3) above). In our experiments as shown in the Supplementary Material, we do encounter some models whose generated medial meshes are not topology preserving. We leave to our future work a theoretically-sound answer to question (3) with topology preservation (see Sec.~\ref{sec:limitations}). Fig.~\ref{fig:teaser} shows a gallery of our feature-preserving medial meshes computed from input CAD meshes with sharp features. 

The contributions of this paper can be summarized as follows: 
\begin{itemize}
    \item First, we formulate an energy optimization framework (in Sec.~\ref{sec:update}) for all medial spheres of different types, to update their positions and radii as close as possible to ground truth, based on the tangential information derived from their surface RPCs.
    \item Second, we present a complete RPD-based framework (in Sec.~\ref{sec:init_mm}) for computing 3D medial mesh while preserving medial features for any input 3D mesh surfaces. We show that the medial mesh derived from RRT (as dual of RPD) could be refined with our geometry-guided thinning algorithm (in Sec.~\ref{sec:refine}), in order to keep the thinness property of the medial axis.
    \item Third, we propose three feature preservation strategies (in Sec.~\ref{sec:feature_preservation}) for preserving external edge features, external corner features, and internal features, under the guidance of surface RPCs. Our method not only preserve external features that none of existing methods can work, but also requires fewer medial sphere samples to generate high quality internal features compared with existing methods.
\end{itemize}

\section{Related Works}

In this section, we shall review the representative approaches for MAT computation. For more extensive discussion on medial axes and other medial representations, we refer readers to those survey articles~\cite{siddiqi2008medial, tagliasacchi20163d}. 

Computing 3D \emph{exact MAT} of polyhedra with feature preservation was explored by earlier algorithms \cite{milenkovic1993robust,sherbrooke1996algorithm,culver2004exact} using seam tracing. Due to the complexity of these exact computation algorithms, they mainly aim for computing medial axes of a simple class of shapes, \eg polyhedra composed of up to about 20 faces, which is impractical for more complicated models. Note that all the models shown in this paper have over tens of thousands of faces. Thus a significant branch of later researches resort to compute an \emph{approximated MAT} (instead of exact computation) in order for them to be suitable for real-world applications~\cite{amenta2001power,amenta2001powerjournal,dey2002approximate,dey2004approximating,pizer2003multiscale,chazal2005lambda,miklos2010discrete,saha2016survey,sobiecki2014comparison}.

As medial axis is notorious for its sensitivity to boundary perturbations, a closely related problem is \textit{MAT simplification} to identify significant and stable parts of the medial axis, and we refer reader to recent excellent works \cite{faraj2013progressive,li2015q,yan2016erosion,pan2019q, dou2021coverage}. While all existing MAT simplification methods are designed for smooth shapes, we believe there could be new MAT simplification methods for non-smooth shapes in the future, by following our feature preservation framework of this paper. It is worth noting that many MAT simplification methods require an initial approximation of MAT in order to prune the noisy branches of the medial axis. The method presented in this paper can provide an initial feature-preserved MAT for any future MAT simplification methods for CAD models.

For 2D smooth shape, it has been proved that the subset of Voronoi diagram of boundary samples provides a structure topologically and geometrically converging to the medial axis \cite{brandt1992continuous}. Unfortunately, this approximation does not hold for 3D smooth shapes due to the existence of ``slivers'' which are tetrahedra with small volume in Delaunay triangulation of boundary samples. This leads to their Voronoi vertices (centers of circumscribing spheres for sliver tetrahedra) very close to boundary but far away from the medial axis \cite{amenta2001powerjournal}. 

\textit{Angle-based filtering methods} are approaches to filter the Voronoi diagram of obtained boundary samples or other derivative structures to approximate the medial axis. Amenta et al. \shortcite{amenta2001power} proposed ``poles'' of Voronoi diagram and show that the \textit{power shape} converges to the medial axis as the sampling density increases. Several other methods \cite{brandt1992continuous,dey2002approximate,dey2004approximating} consider a subset of Voronoi diagram of boundary samples that satisfy an angle criteria given a user-specified threshold. These methods are known to be difficult to preserve the topology of the input shape as the filtered subset tends to have many holes and isolated elements.

\textit{$\lambda$-medial axis methods}  \cite{pizer2003multiscale,chazal2005lambda} use the radius of the closest medial sphere as filtering criteria and discard a medial sphere if its radius is smaller than a given threshold $\lambda$. As a result, the medial axis consists of medial spheres such that the smallest enclosing sphere of the nearest boundary sample set has a radius equals or larger than $\lambda$.

\textit{Voxel-based methods} belong to an completely different category that approximate the medial axis of voxel shape by selecting a subset of voxels that share similar properties as medial axis \cite{saha2016survey,sobiecki2014comparison}. The state-of-the-art method Voxel Core \cite{yan2018voxel} can well approximate the medial axis of any smooth shape while guaranteeing the topological correctness of the generated medial axis, given a voxelization of the shape at sufficiently high resolution. A common drawback of these methods is that they require fine voxel resolutions hence have high computational cost in order to achieve a comparable geometric accuracy as sampling-based methods. Voxel Core requires an additional $\lambda$-pruning which shrinks the medial axis while removing ill-posed structure. This pruning operation makes the medial mesh incomplete especially for non-smooth regions like sharp edges or corners.

None of the above MAT approximation methods, however, consider handling 3D shapes with non-smooth regions such as sharp edges and corners (as we call them \emph{external features}), and none of them consider preserving the internal features of medial axis. To the best of our knowledge, only Dey et al. proposed a remedial method CAD\_MEDIAL \cite{dey2003approximate} which extends their previous method MEDIAL \cite{dey2002approximate} to complete the structure of medial axis with external features. But the sampling condition proposed in CAD\_MEDIAL near sharp edges is too impractical to be achieved, leading to failures to preserve external features in many shapes. Comparing with these methods, our method for approximating the medial axis is the first one that preserves not only external features but also internal features of the medial axis. 


\begin{figure*}[t]
    \centering
    \includegraphics[width=\linewidth]{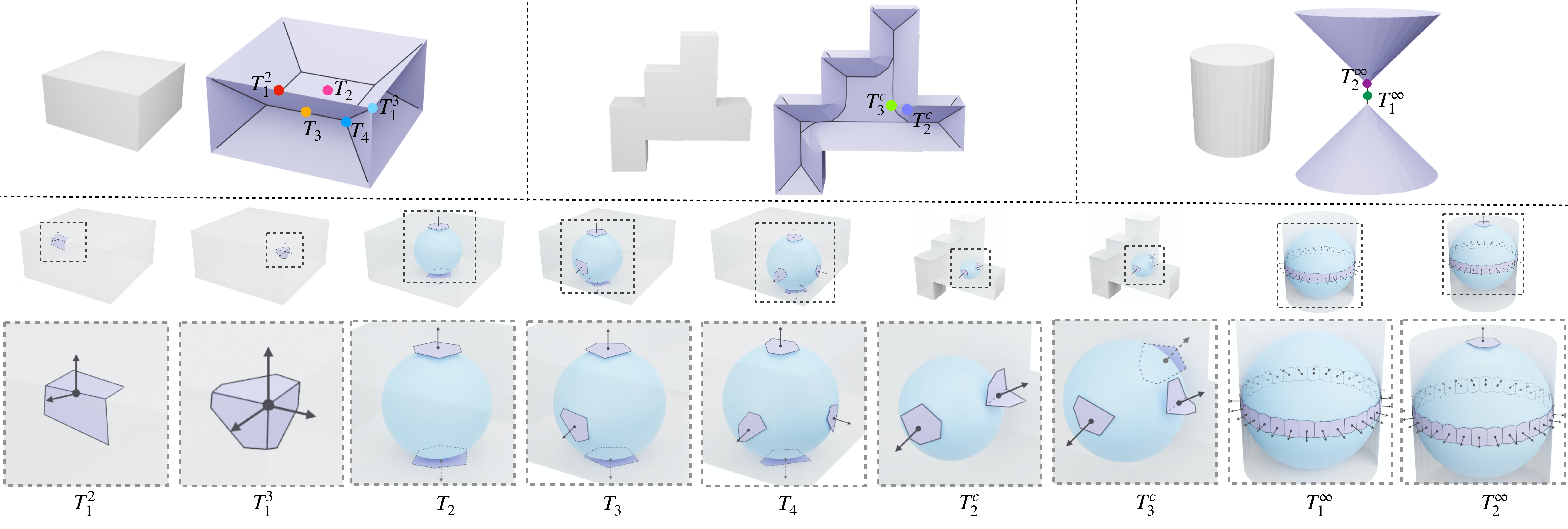}
    \vspace{-15pt}
    \caption{Examples of different classes of medial spheres in the context of their connected components (CCs) of restricted power cell (RPC). For a medial sphere, the subscript of its class type represents the number of CCs of its RPC, and the superscript represents the number of surface normals for the contact points in each CC. Note that we use the simplified notations here by removing all appearance of ``1'' from the superscript, \ie $T_3^{1,1,1}$ is simply written as $T_3$.}
    \label{fig:medial_types}
\end{figure*}

\section{Preliminaries}

\subsection{Medial Axis and Medial Mesh}
\label{sec:pre_ma}

Given a closed, oriented, and bounded shape $\model \in \rspace^3$, the \textit{medial axis} $\ma$ is defined as the locus of centers of spheres that are tangent to two or more points on the boundary of $\model$, or $\bmodel$, without containing any other boundary points in its interior. 

The \textit{medial axis transform} (MAT) is formed by the medial axis $\ma$ together with its radius function. To approximate the MAT of a 3D shape $\model$, we use a non-manifold \textit{medial mesh} $\mmesh$ consisting of triangles and edges. Each vertex of $\mmesh$ represents a medial sphere $ \msphere = (\mcenter, r)$, where $\mcenter \in \rspace^3$ is the sphere center and $r \in \rspace $ is its radius. MAT can be used to reconstruct the surface through the union of enveloping volumes of its medial primitives~\cite{li2015q}. For example, the enveloping volume of an edge of the medial mesh is called a \textit{medial cone}, which is the linear interpolation of two spheres $e_{ij} = t \msphere_i + (1-t) \msphere_j$, $t \in [0, 1]$. The enveloping volume of a triangle face $f_{ijk}$ of the medial mesh is called a \textit{medial slab} that is the linear interpolation of three spheres $\msphere_i$, $\msphere_j$, and $\msphere_k$.

In this paper, we assume the boundary surface $\bmodel$ is provided as a watertight and manifold 3D triangular mesh, with those sharp feature edges and corners pre-labeled on the mesh. We do not have any other assumption about these sharp features, \eg how small a dihedral angle is allowed to be on a sharp edge, etc. In fact, the case when multiple sharp edges (convex and/or concave) meet at a corner is particularly challenging, \eg two adjacent sharp edges forming a small angle around a corner. Our feature preservation strategy is designed to handle all these challenging cases as discussed in Sec.~\ref{sec:feature_preservation}.

\subsection{Restricted Power Diagram (RPD)}
\label{sec:pre_RPD}

The Voronoi diagram of a set of generators $\{ \mcenter_i \}_{i=1}^{n}$ is a partition of the domain $\Omega \subset \rspace^d$ into a set of cells. Each cell $\vorcell_i$ consists of the points $\anypoint \in \Omega$ closest to a particular generator $\mcenter_i$:
\begin{equation}
    \vorcell_i: \{ \anypoint \in \Omega | ||\anypoint - \mcenter_i|| \leq ||\anypoint - \mcenter_j||, j \neq i \}.
\end{equation}
The vertices of these cells are called \textit{Voronoi vertices}. The \textit{Voronoi ball} in $\rspace^3$ centered at a Voronoi vertex has at least $4$ generators on its boundary, and no generator in its interior \cite{amenta2001power}.

Power diagram \cite{aurenhammer1987power} is a generalization of Voronoi diagram by weighting the given generators, and coincides with Voronoi diagram in the special case that all points have equal weights. Given a set of weighted generators $\{ \msphere_i=(\mcenter_i, r_i) \}_{i=1}^{n}$, a \textit{power cell} $\powercell_i$ is defined as:
\begin{equation}
     \powercell_i: \{ \anypoint \in \Omega | d_{pow}(\anypoint, \msphere_i) \leq d_{pow}(\anypoint, \msphere_j), j \neq i \},
\end{equation}
where $d_{pow}(\anypoint, \msphere_i) = ||\anypoint - \mcenter_i||^2 - r_i^2$ is the power distance between any point $\anypoint$ and the weighted generator $(\mcenter_i, r_i)$. 

A power diagram restricted within a bounded shape $\model \subset \rspace^n$ is called a \textit{restricted power diagram} (RPD) $\mathcal{R}$, where all power cells overlap with the shape $\model $. The RPD consists of a set of \textit{restricted power cells} (RPC) $\{ \rpc_i \}_{i=1}^n$. Each cell $\rpc_i$ is the restriction of the power cell $\powercell_i$ of the weighted generator $(\mcenter_i, r_i)$ within $\model $: 
\begin{equation}
\rpc_i = \powercell_i \cap \model. 
\end{equation}
The surface RPC is the restriction of the power cell within the boundary surface of the shape, denoted as: 
\begin{equation}
\partial \rpc_i = \powercell_i \cap \bmodel.
\end{equation}
Robust and exact computation of surface RPD on a surface represented as triangle mesh is a non-trivial task. Our implementation extends Yan et al.'s algorithm \shortcite{yan2009isotropic} to compute them robustly and exactly, by replacing the metric from Euclidean to power distance.

\begin{figure*}[h]
    \centering
    \includegraphics[width=\linewidth]{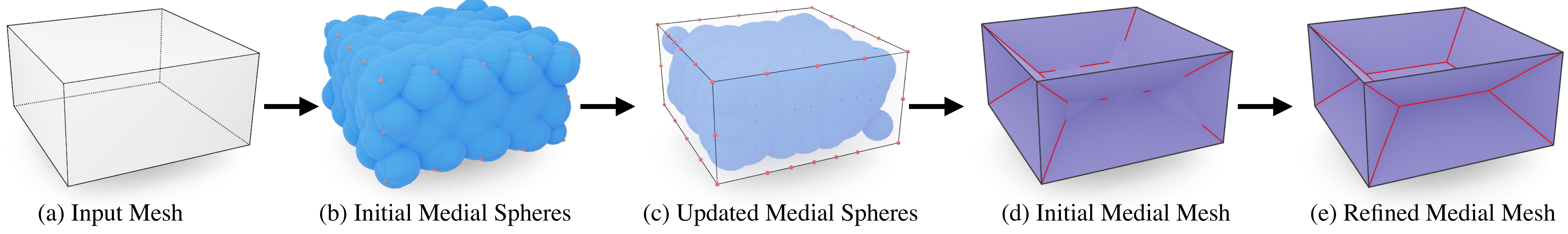}
    \vspace{-15pt}
    \caption{The pipeline of our algorithm. Given a closed manifold 3D mesh surface (a) with sharp edges and corners pre-detected (marked in black lines), the initial medial spheres (b) are generated including zero-radius spheres (shown as red points) placed on convex sharp features and inner Voronoi balls (shown as blue spheres) of surface samples, as described in Sec.~\ref{sec:initial}. Inner Voronoi balls protrudes from the surface are then updated and pushed to be tangential to the surface (c), as described in Sec.~\ref{sec:update}. An initial medial mesh (d) is then constructed based on RRT, and internal features (shown as red lines) are selected using a seam tracing algorithm, as described in Sec.~\ref{sec:init_mm}. A refined medial mesh (d) with higher quality can be obtained from feature preservation strategies and a thinning process, as described in Sec.~\ref{sec:refine}.}
    \label{fig:overview}
\end{figure*}

\subsection{Classification of Medial Spheres}
\label{sec:classification}

For a medial sphere $\msphere_i = (\mcenter_i, r_i)$ that is tangent to the boundary surface at two or more points, its surface RPC $\brpc_i$ corresponds to an RPD constructed on the surface with respect to $\msphere_i$, which consists of a set of \textit{Connected Components (CCs)} of surface regions, each of which contains one or more tangential surface points. For a smooth object, the medial spheres are organized into a small amount of classes \cite{giblin2004formal}. Since we are targeting for non-smooth surfaces, we count the number of surface normals of those tangential points. We use our own notation $T_k^{n_1,...,n_k}$ for the classification of a medial sphere $\msphere_i$. Here the subscript $k$ represents the number of CCs of its RPC, and the superscript $n_1,...,n_k$ represents the number of surface normals for the tangential points in each CC. 

For example, medial spheres of type $T_2^{1,1}$ lie on 2-manifold sheets, which are tangent to $\bmodel$ at exactly two distinct CCs and each CC contains only one tangential point (with its corresponding normal). Since this is the most ordinary case, to simplify the notations we remove all appearances of $1$ from the superscript, \textit{i.e}, $T_2^{1,1}$ will be simply denoted as $T_2$. For most 3D shapes, the majority of medial spheres lies on type $T_2$ of \textit{sheets}; the intersection of three or more local sheets forms a \textit{seam} of $T_3$ spheres; and seams of type $T_3$ could intersect at a \textit{junction} sphere of type $T_4$. Fig.~\ref{fig:medial_types} gives examples of these different classes. 



In this paper we are handling non-smooth surfaces that could contain sharp edges and corners (either convex or concave). The tangential surface contact point on these sharp features does not have a unique and unambiguous normal. We define a \textit{convex sharp edge} and \textit{concave sharp edge} of input mesh $\bmodel$ as an edge subtending a dihedral angle less than $\pi - \sharpangle$ and more than $\pi + \sharpangle$ respectively \cite{abdelkader2020vorocrust}, where $\sharpangle < \frac{\pi}{2}$ is an angle threshold used to bound the approximation error. Note that $\sharpangle$ is a user-defined variable and users can also mark sharp features manually. A vertex of $\bmodel$ located on more than two sharp edges is defined as a \textit{corner}.
\begin{itemize}
\item On a \emph{convex sharp feature}, the medial spheres are of zero radii. The RPC of such zero-radius sphere have only one CC, but there could be two or three unambiguous normals associated with such feature point: sharp edges have two normals and sharp corners have at least three. Thus we denote such zero-radius medial spheres as $T_1^2$ (on a convex edge) and $T_1^u$ ($u\geq3$, on a corner). Note that a corner could be formed by a combination of both convex and concave sharp edges (Sec.~\ref{sec:extf_add_corner} discusses these different cases and the mechanism to preserve the corner). For a corner that is formed by purely concave feature edges, there is no medial axis passing through it, thus the mechanism to handle the medial spheres adjacent to it is the same as the other concave features mentioned below.
\item On a \emph{concave sharp feature}, the medial spheres tangential to such features are not zero-radius. In fact, the normal direction of such tangential surface contacts are ambiguous (Sec.~\ref{sec:initial} gives detailed information of such medial spheres). Thus we use superscript $c$ for annotating a CC that contains a tangential contact point on any concave external feature. See the examples of $T_2^{c}$ and $T_3^{c}$ in Fig.~\ref{fig:medial_types}. Note that a medial sphere could be tangential to more than one concave feature edges on different CCs. Here we do not make distinct notations for them, as their sphere computation mechanism is similar to the case of single concave edge (see Sec.~\ref{sec:update} for the details).
\end{itemize}
The CC containing an infinite number of tangential contact points will be annotated with superscript $\infty$, \textit{e.g}, for a cylinder shape, the CC of a medial sphere could contain the whole circular region that has an infinite number of tangential contact points with the sphere. In practice, since our input surface is represented as a triangle mesh, the number of tangential points for such a sphere is finite. Thus the superscript $\infty$ is only for notational purpose. See the examples of $T_1^{\infty}$ and $T_2^{\infty}$ in Fig.~\ref{fig:medial_types}. They could be identified whenever a CC has more than one tangential point with diverse normals.

The \textit{external features} of MAT include those convex sharp edges and their associated corners, and the \textit{internal features} of MAT include those internal spheres located on \textit{seams} (\eg types of $T_3$ and $T_3^{c}$) as well as \textit{junctions} (\eg types of $T_4$, $T_4^{c}$, and $T_2^{\infty}$). Note that the examples of classes given in Fig.~\ref{fig:medial_types} are not meant to be complete. For example, there could be a corner denoted as $T_1^{6}$ formed by three convex edges and three concave edges, and there could be a junction sphere $T_4^{c}$ that is tangential to two concave sharp edges and two other regular CCs, just to name a few. But notation-wise, all medial spheres can be represented with our CC-based classification. With this classification, the following sections discuss how we can compute these medial spheres based on such CC information, and how we can connect them to form a medial mesh while preserving both external and internal features of MAT.

\section{The Computational Pipeline}
\label{sec:computational_pipeline}

Our medial mesh computational pipeline consists of four major steps. Given a closed, manifold triangular mesh with sharp features pre-detected, the first step (Sec.~\ref{sec:initial}) is to initialize the medial spheres. This includes placing sphere candidates on both non-feature and feature regions. The sphere candidates on non-feature regions are inner Voronoi balls generated with Delaunay Triangulation of surface samples. Special attention needs to be paid on initializing spheres tangential to concave sharp edges. The second step (Sec.~\ref{sec:update}) updates those inner sphere candidates as close as possible to their ground-truth positions and radii using our sphere updating strategy. Then in the third step (Sec.~\ref{sec:init_mm}) we construct an initial medial mesh from these updated sphere candidates using the restricted regular triangulation (RRT) which is dual to RPD. Initial internal features could be detected through our seam tracing algorithm. In the fourth step (Sec.~\ref{sec:refine}), the initial medial mesh is further refined through our internal feature preservation strategy and a thinning process to pursue the thinness property of medial mesh.
Fig.~\ref{fig:overview} shows an illustration of our computational pipeline.

\subsection{Medial Sphere Initialization}
\label{sec:initial}

\paragraph{Non-feature Spheres:} Our method starts with the generation of inner sphere candidates in those non-feature regions. A well-known technique is to use Voronoi balls (see Sec.~\ref{sec:pre_RPD}) inside the shape $\model$ generated from some sampling points $\sample$ on the boundary surface $\bmodel$, and use the fast winding number \cite{Barill:FW:2018} to keep only those inner Voronoi balls.
Our assumption about the density of sample set $\sample$ around smooth regions follows Amenta et al.'s work \shortcite{amenta2001power}. We use the local feature size function $\lfs: \bmodel \rightarrow \rspace$ defined as the minimum Euclidean distance from a sample $\mathbf{x} \in \bmodel$ to the medial axis $\ma$. A sample set $\sample$ is an \textit{r-sample} if any sample $\mathbf{x} \in \sample$ has a neighboring sample within its $r\lfs$ distance. Similar to existing Voronoi-based algorithms \cite{amenta2001power,dey2002approximate}, we require $r \leq 0.6$ for sufficiently approximating $\bmodel$ so that the generated Voronoi diagram can capture key information about shapes. Since the LFS around the convex sharp edges and corners are converging to zero, we stop the surface sampling that are within an $\eta$ distance to these convex external features, where $\eta$ is a distance threshold to avoid the density explosion to infinity around these external features.

\paragraph{Zero-radius Feature Spheres:}
To avoid dealing with infinity density around convex sharp features, we apply an adaptive medial re-sampling strategy. We first sample zero-radius medial spheres on those pre-detected convex sharp edges and corners, with the initial density same as its nearby surface samples that are $\eta$-distance away. 
After all those non-feature medial spheres are updated in the second step (Sec.~\ref{sec:update}), we might insert new zero-radius spheres to preserve the external feature. Specifically, for every non-feature medial sphere $\msphere_x$ whose RPC is neighbor to those of zero-radius spheres on external features, we recursively check and add new zero-radius feature sphere if $\msphere_x$ breaks the connection of external features. Please refer to Sec.~\ref{sec:extf_add_se} in more detail. Note that in this first step we only add zero-radius feature spheres using the initial density, while adaptive sphere insertion is recalled once the neighboring non-feature medial spheres are updated in the second step. Similarly, simply sampling zero-radius medial spheres on sharp corners is not enough for keeping the connectivity of medial axis. To preserve corner feature, we analyze the structure around corners and sample new medial spheres nearby to complete the medial structure. Please refer to Sec.~\ref{sec:extf_add_corner} for detailed explanation.

\begin{figure}[h]
    \centering
    \includegraphics[width=0.8\linewidth]{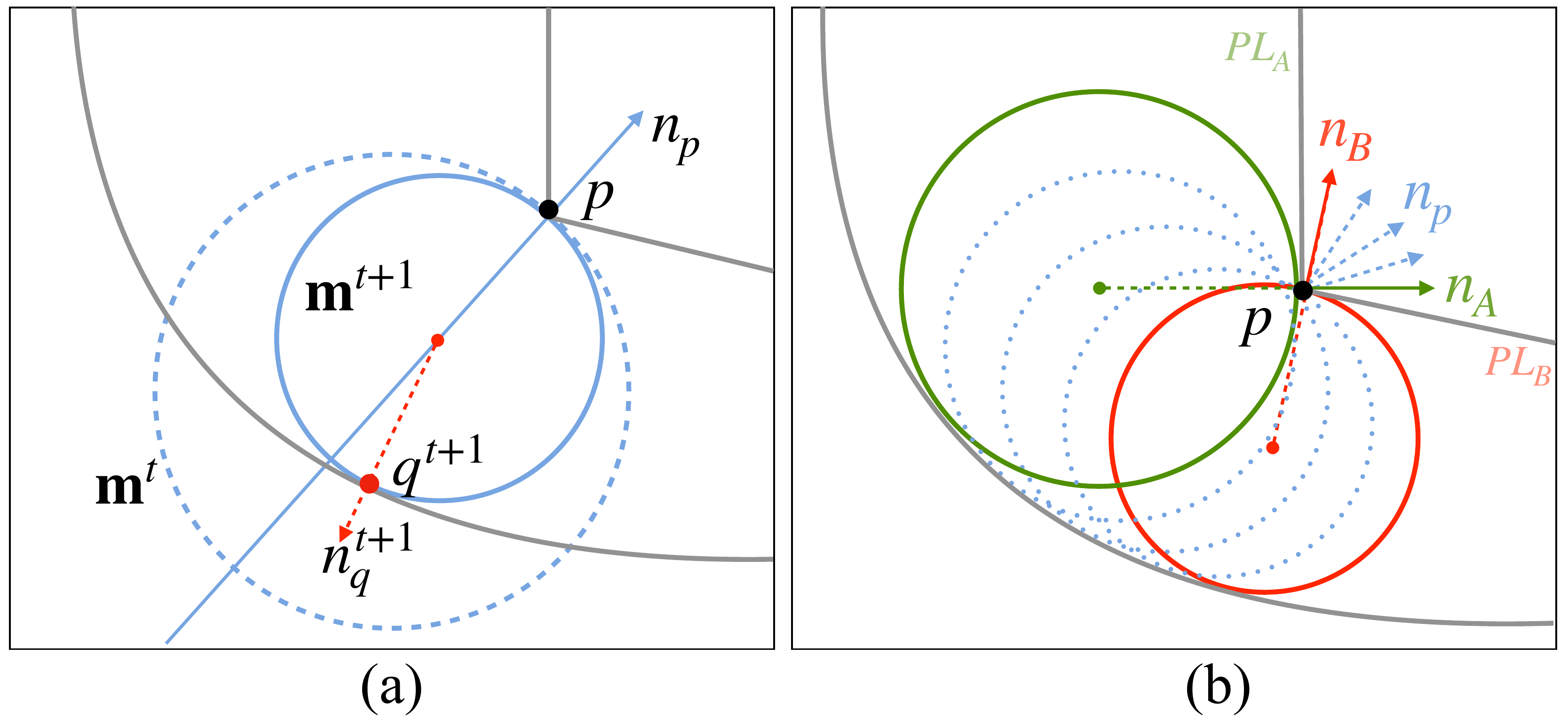}
    \caption{Left: The \textit{sphere-shrinking} algorithm \cite{ma20123d} on concave regions in 2D. Right: Medial sphere initialization around concave region in 2D. For every pin-point $\tanpoint$ sampled on concave sharp edge, we sample new normal $\tannormal_{\tanpoint}$ between two normals $\tannormal_A$ and $\tannormal_B$ of planes adjacent to the concave edge, and initialize medial spheres using the sphere-shrinking algorithm. The denser the sampled pairs of $\{\tanpoint,\tannormal_{\tanpoint}\}$, the smoother the final medial mesh around the concave edge we can get.}
    \label{fig:sphere_init}
\end{figure}

\paragraph{Spheres Tangential to Concave Sharp Features:}
Even though concave sharp edges are not external features, a sharp change of surface normals around them leads to a smooth transition on medial axis (see Fig.~\ref{fig:sphere_init} (b)), which requires dense samples of medial spheres around them. Since all concave sharp edges are pre-detected, we can sample dense medial spheres using \textit{sphere-shrinking} algorithm \cite{ma20123d}. 
We briefly introduce the algorithm as shown in Fig.~\ref{fig:sphere_init} (a). For each pin-point $\tanpoint$ with corresponding normal $\tannormal_{\tanpoint}$, a sphere $\msphere^0$ with large radius $r^0$ and center $\mcenter^0 = \tanpoint - r^0 \tannormal_{\tanpoint}$ is initialized. This large sphere is then iteratively shrunk to approximate the medial sphere with two tangent points. For each iteration $t+1$, a new sphere $\msphere^{t+1}$ is found by performing a nearest point query from $\mcenter^t$ to the surface $\bmodel$ excluding point $\tanpoint$. The resulting nearest point $\mathbf{q}^{t+1}$ together with $\tanpoint$ is then used to compute the new sphere $\msphere^{t+1}$. The iteration stops when the sphere $\msphere^{t+1}$ is tangential on these two points and there is no other surface points closer to the sphere center. Our sampling strategy on concave sharp edges works as follows (see Fig.~\ref{fig:sphere_init} (b) for a 2D illustration): for each pre-detected concave edge $\ccl$, we sample pin-points $\tanpoint$ densely on $\ccl$ and sample multiple normals $\tannormal_{\tanpoint}$ in between two normals $\mathbf{n}_A$ and $\mathbf{n}_B$ of adjacent planes of the concave edge $\ccl$. For each pair of ($\tanpoint$, $\tannormal_{\tanpoint}$), we initialize a sphere that is large enough and then apply sphere-shrinking algorithm. The sphere-shrinking algorithm is guaranteed to converge \cite{ma20123d} and it typically converges within a couple of iterations. It perfectly fits medial spheres with two tangential points, which are of types $T_2$ or $T_2^{c}$ on sheets of the medial axis.

After the sphere initialization, however, inner Voronoi balls initialized for smooth regions are circumscribed over the surface sampling points in nature, making themselves often protrude the surface $\bmodel$. Moreover, Voronoi-based generation in $\rspace^3$ is notorious for containing a large amount of Voronoi balls called \emph{spikes} that are very close to the surface and far from the medial axis \cite{amenta2001power}. Even though medial spheres sampled on pre-detected concave edges are already at their ideal position using the sphere-shrinking algorithm, the internal features of MAT such as seams and junctions may not be fully represented. This is because the sphere-shrinking algorithm cannot sample  spheres tangential to more than two surface points. Thus in the next step, these initial Voronoi spheres will be updated, and spikes will be removed. 

\subsection{Medial Sphere Update}
\label{sec:update}

Given the initial medial spheres estimated from inner Voronoi balls, we apply local operations to update their positions and radii as close as possible to the ground truth. Ideally we expect all medial spheres to be tangent to at least two points on surface $\bmodel$ without protrusion. 

For inner Voronoi balls initialized with two tangent points, \ie type $T_2$ or $T_2^{c}$ on medial sheets, we use sphere-shrinking algorithm \cite{ma20123d} as described above to update them. Although this algorithm is highly efficient, it cannot handle spheres with more than two tangent points, \ie $T_3$ and $T_3^{c}$ on medial seams, or $T_4$ and $T_4^{c}$ on medial junctions. The problem is more challenging than the case of two tangent points, because it is very difficult to find an exact pin-point on either the seam or junction to determine the sphere. In this paper we formulate the updating algorithm for spheres with any number of tangent points as a continuous optimization problem, and name it as \emph{multi-tangent sphere optimization} algorithm. 

\begin{figure}[h]
    \centering
    \includegraphics[width=0.85\linewidth]{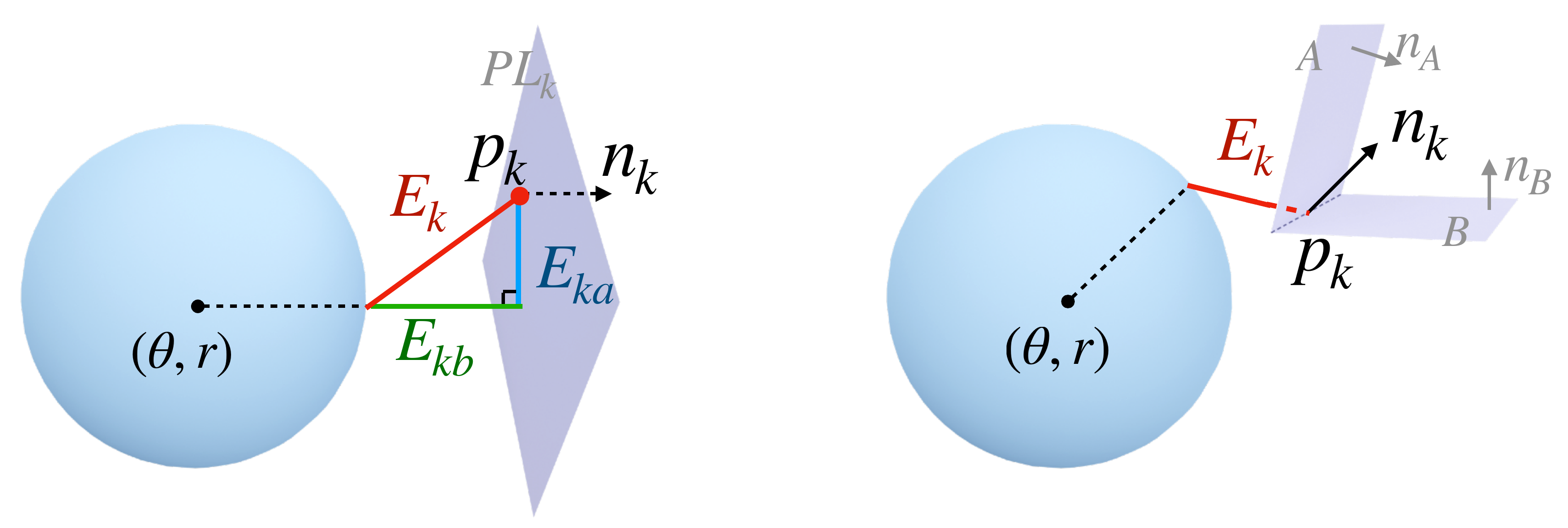}
    \caption{Illustration of energy terms for our multi-tangent sphere optimization. Left: Energy term $\energy_k$ as the sum of $\energy_{ka}$ and $\energy_{kb}$ when the tangential point $\tanpoint_k$ is on a plane $\tanpl_k$. Right: energy term $\energy_k$ when the tangential point $\tanpoint_k$ is on a concave sharp edge $\ccl_k$.}
    \label{fig:sphere_update}
\end{figure}

Our multi-tangent sphere optimization is inspired by the property of surface RPC. Each $CC$ of sphere $\msphere$ defines at least one tangent plane or tangent concave edge of $\msphere$ with their tangential points $\tanpoint_k$ and normals $\tannormal_k$, where $k$ is the index of all tangential points belonging to $\msphere$. In this problem, not only the sphere $\msphere=(\mcenter,r)$ needs to be solved, but also the exact positions and normals ($\tanpoint_k$, $\tannormal_k$) of all tangential points are unknown.
We define the following quadratic energy for our multi-tangent sphere optimization: 
\begin{equation}
\begin{split}
     &\energy(\mcenter,r,\{\tanpoint_k, \tannormal_k\}_{k=1}^{N}) = \sum_{k=1}^{N}  \energy_k, \\
    &\energy_k = || \mcenter + r \tannormal_k - \tanpoint_k ||^2.
\label{eq:energy}
\end{split}
\end{equation}
Here $N$ is the total number of tangential points for this sphere, and $\energy_k$ is the energy term defined using the tangent pair ($\tanpoint_k$, $\tannormal_k$). Note that tangential point $\tanpoint_k$ can be either on a plane (Fig.~\ref{fig:sphere_update} (a)) or on a concave sharp edge (Fig.~\ref{fig:sphere_update} (b)). We define $\energy_k$ as the squared Euclidean distance from $\tanpoint_k$ to the expected point of tangency on sphere. This energy is zero when the sphere is exactly tangential to the surface at point $\tanpoint_k$.

As shown in Fig.~\ref{fig:sphere_update} (a), we can split the energy term $\energy_k$ as the sum of two sub-terms $\energy_{ka}$ and $\energy_{kb}$: 
\begin{equation}
\begin{split}
\energy_{kb} &= || (\tanpoint_k - \mcenter)^{\top} \tannormal_k  - r||^2, \\
\energy_{ka} &= \energy_{k} - \energy_{kb}.
\end{split}
\end{equation}
Intuitively $\energy_{kb}$ is the squared distance from sphere $\msphere$ to plane $\tanpl_k$ along the direction of normal $\tannormal_k$, and $\energy_{ka}$ is the squared distance on the plane. We modify the energy in Eq.~\eqref{eq:energy} by incorporating two weights $\lambda_a$ and $\lambda_b$ to balance the significance of these two sub-terms:
\begin{equation}
     \energy(\mcenter,r,\{\tanpoint_k, \tannormal_k\}_{k=1}^{N}) = \sum_{k=1}^{N}  \lambda_a\energy_{ka}+\lambda_b\energy_{kb}.
\label{eq:energy_updated}
\end{equation}
Inspired by the classic iterative closest point (ICP) algorithm~\cite{CHEN1992ICP}, giving a higher weight $\lambda_b$ to the tangential squared distance $\energy_{kb}$ could potentially speed up the convergence of optimization. We use $\lambda_a = 0.01$ and $\lambda_b = 1$ in all of our experiments.

If $\tanpoint_k$ is on a smooth surface region (Fig.~\ref{fig:sphere_update} (a)), then its normal $\tannormal_k$ is fully determined by its position. If $\tanpoint_k$ is on a concave sharp edge $\ccl_k$ (Fig.~\ref{fig:sphere_update} (b)), then $\tannormal_k$ can be any direction in between two normals $\mathbf{n}_A$ and $\mathbf{n}_B$ of adjacent planes of the concave edge. These two normals give us a bound of possible directions that the medial sphere can be tangent to. 

\begin{figure}[h]
    \centering
    \includegraphics[width=0.9\linewidth]{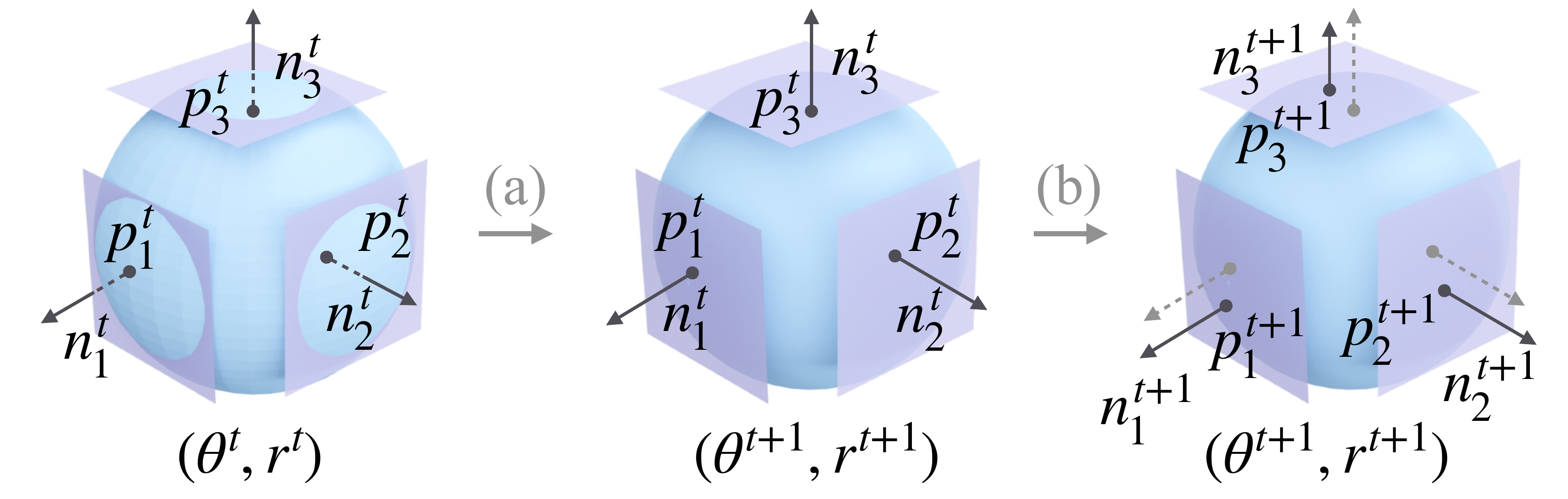}
    \caption{Iterative two-stage optimization for medial spheres. (a) The sphere-updating stage which locks the aggregated tangent pairs ($\tanpoint_k$, $\tannormal_k$), $k=1...N$ and update the medial sphere $(\mcenter, r)$; (b) The tangent-updating stage that fix the previously updated medial sphere then update each tangent pair. Both stages optimize the same energy function in Eq. \eqref{eq:energy_updated}.}
    \label{fig:sphere_update_optimization}
\end{figure}

\begin{figure}[h]
    \centering
    \includegraphics[width=\linewidth]{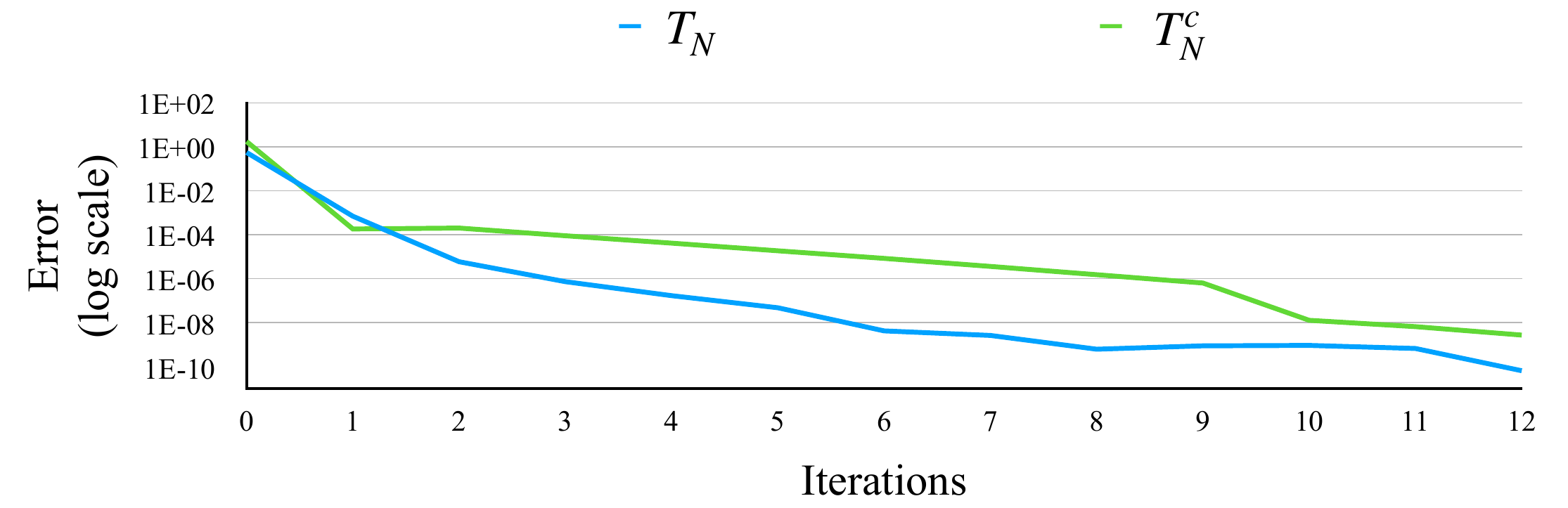}
    \caption{Convergence of the sphere updating optimization of Eq. \eqref{eq:energy_updated} through iterations (in log scale for better illustration). The error is measured as the squared distance from the sphere to the surface. We show the average errors collected from all models shown in Fig.~\ref{fig:more_rec} aggregated in two sphere types: $T_N$ and $T_N^c$, where $N > 2$. Note that $T_2$ and $T_2^c$ are not collected since we use sphere-shrinking algorithm \cite{ma20123d} in Sec.~\ref{sec:initial}}.
    \vspace{-5pt}
    \label{fig:sphere_update_spline}
\end{figure}

Since both the sphere $\msphere=(\mcenter,r)$ and the tangent pairs ($\tanpoint_k$, $\tannormal_k$), $k=1...N$ are to be determined, we design our multi-tangent sphere optimization algorithm as an iterative two-stage optimization process. For each new iteration $t+1$, we decompose the optimization into a \emph{sphere-updating stage} and a \emph{tangent-updating stage}:
\begin{itemize}
    \item During the sphere-updating stage (Fig.~\ref{fig:sphere_update_optimization} (a)), we fix all tangent pairs ($\tanpoint_k^{t}$, $\tannormal_k^{t}$), $k=1...N$ from the previous iteration $t$, and update the sphere ($\mcenter^{t+1}$, $r^{t+1}$) by minimizing the energy in Eq.~\eqref{eq:energy_updated}. Since the energy is a quadratic function of $\mcenter^{t+1}$ and $r^{t+1}$, they can be solved by a simple linear equation.
    \item During the tangent-updating stage  (Fig.~\ref{fig:sphere_update_optimization} (b)), we lock the sphere ($\mcenter^{t+1}$, $r^{t+1}$) from the first stage and update all tangent pairs ($\tanpoint_k^{t+1}$,$\tannormal_k^{t+1}$), $k=1...N$. If $\tanpoint_k^{t}$ is on a plane $\tanpl_k$ of the surface, we search the $\numneighbors$-ring neighboring surface triangles around $\tanpoint_k^{t}$ ($\numneighbors=2$ in our experiments), and find a point within these triangles that minimizes the quadratic energy $\energy$ as the new tangent point. As each triangle has a fixed normal, searching inside a triangle is equivalent to solving a linear system of barycentric coordinates. If $\tanpoint_k^{t}$ is on a concave sharp edge $\ccl_k$, the optimal tangent point $\tanpoint_k^{t+1}$ is simply the projection of the sphere center $\mcenter^{t+1}$ onto the concave sharp edge, and the optimal normal $\tannormal_k^{t+1}$ is the normalized direction of $\tanpoint_k^{t+1}-\mcenter^{t+1}$. Note that $\tannormal_k^{t+1}$ will be clamped to boundary normal once out of range of $\mathbf{n}_A$ and $\mathbf{n}_B$.
\end{itemize}

Since both stages decrease the quadratic energy $\energy$ in Eq.~\eqref{eq:energy_updated}, the optimization is converging through iterations. We visualize the convergence of our iterative optimization scheme in Fig.~\ref{fig:sphere_update_spline}. Note that the iteration process does not require a re-calculation of RPD. All inner Voronoi balls will be updated iteratively until their energy functions $E$ are smaller than a threshold $\epsilon$ ($10^{-4}$ in our experiment). A converging energy larger than $\epsilon$ indicates that such a sphere does not exist to be tangential to the set of given planes and/or concave edges, and it will be removed consequently. 
This energy-checking will become very useful in our sphere-insertion operations, \eg for external corner preservation described in Sec.~\ref{sec:extf_add_corner}.

\begin{wrapfigure}{r}{0.1\textwidth}
  \centering
  \vspace{-5pt}
  \begin{center}
    \includegraphics[width=0.1\textwidth]{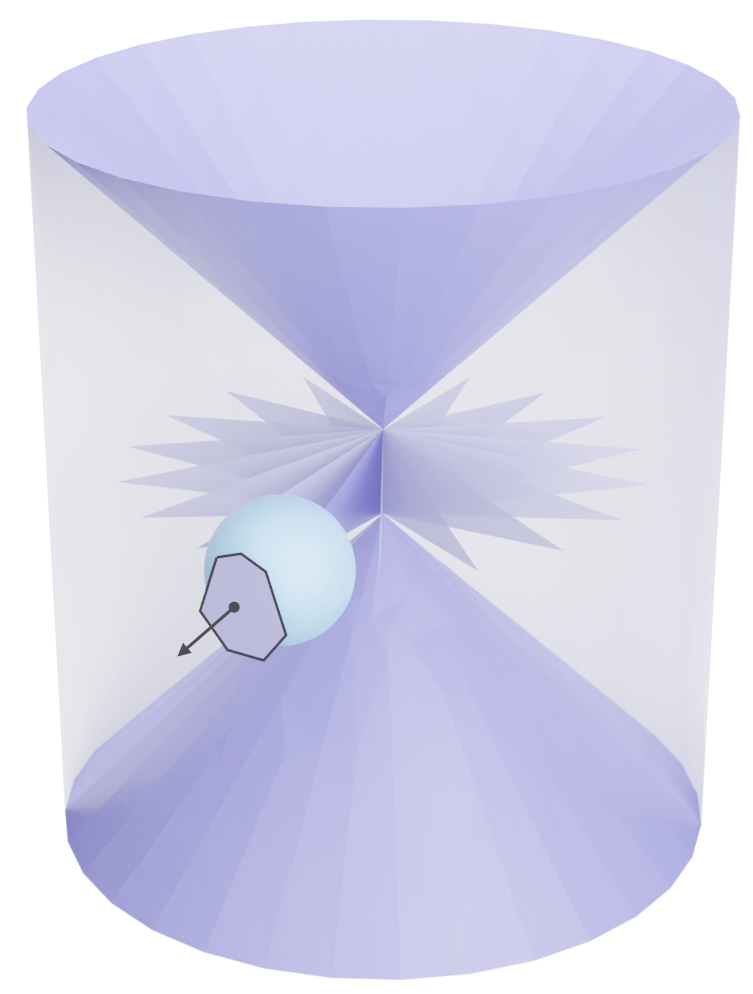}
  \end{center}
  \vspace{-5pt}
\end{wrapfigure}
It is worth noting that spheres of type $T_1$ are normally \textit{spikes}, because they only touch one side of the surface and has only one tangent point and normal (see the right inset figure). Even though our method is not specifically designed for pruning spikes, a by-product of our computed RPC is to robustly detect this type of medial spheres and remove them.

\subsection{Medial Mesh Initialization}
\label{sec:init_mm}

After the sphere update, there are some non-feature spheres that break the connectivity of nearby external features (details in Sec.~\ref{sec:extf_add_se}). We will re-sample zero-radius feature spheres in those cases. Once all spheres (including zero-radius spheres and update inner Voronoi balls) are at their ideal positions (Fig.~\ref{fig:overview} (c)), we build their connectivity to form a structured medial mesh. Our topological connection method is inspired by the \textit{power shape} \cite{amenta2001powerjournal} which is a subset of the \textit{regular triangulation} (RT) dual to the \textit{power diagram} (PD). To construct the medial mesh $\mmesh$, we compute the \textit{restricted regular triangulation} (RRT), by selecting a subset of simplices in RT, whose dual elements in PD have non-empty intersections with the input shape $\model$. Specifically, we check the RT simplicies in the order of tetrahedra, triangles, and edges: 
\begin{itemize}
    \item The dual of an RT tetrahedron is a vertex in PD. If this dual vertex is inside the shape $\model$, then we keep this tetrahedron together with all of its triangles and edges in the medial mesh.
    \item If the dual vertex of the RT tetrahedron is outside the shape, then we check all of its four triangles. The dual of an RT triangle is an edge segment in PD. If this dual edge segment has any intersection with the shape $\model$, then we keep this triangle together with all of its edges in the medial mesh.
    \item If the dual edge segment of the RT triangle is outside the shape, then we check all of its three edges. The dual of an RT edge is a polygonal face in PD. If there is any intersection between the dual polygonal face with the shape $\model$, then we keep this RT edge in the medial mesh. 
\end{itemize}
Note that in this process we only need to compute RT as dual of PD and compute its restriction to the shape $\model$ by checking the intersection between its dual edge segment or polygon with input surface $\bmodel$. In this way we do not need to compute the volumetric RPD which requires a tetrahedralization of the input shape $\model$ and the cutting of those tetrahedra with PD. It is worth mentioning that, similar to power shape~\cite{amenta2001powerjournal}, our initial medial mesh $\mmesh$ generally contains some flat but solid tetrahedra. However, the medial axis of a three-dimensional shape $\model$ should be a collection of two-dimensional sheets, \ie they should be \emph{thin} without any solids. All these tetrahedra in the initial medial mesh will be pruned in our thinning process in the fourth step (Sec.~\ref{sec:refine}).

Once the initial medial mesh is constructed, we can trace the internal features using a seam tracing algorithm similar to Culver et al.'s method~\shortcite{culver2004exact}. 
We first classify initial medial spheres based on their CCs and tangent points, and identify those medial spheres on seams (\ie type $T_3$ or $T_3^{c}$). Our seam tracing algorithm starts from any seam sphere and expand to its adjacent feature spheres. The detailed seam tracing algorithm is provided as Alg.~1 in Supplementary Material.
In this way we can detect an initial set of internal features (Fig.~\ref{fig:overview} (c) red lines) which should be further refined in the next step.

\subsection{Medial Mesh Refinement}
\label{sec:refine}

\begin{figure}[h]
    \centering
    \includegraphics[width=\linewidth]{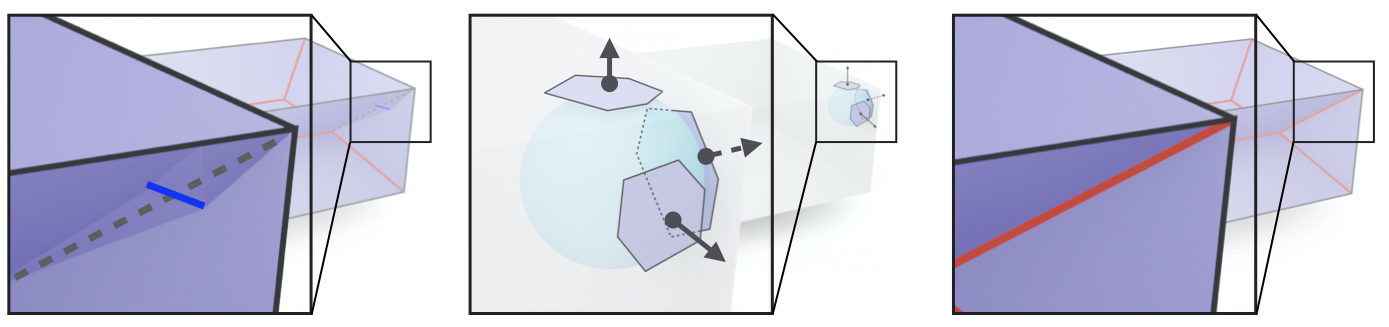}
    \caption{Left: Ill-posed connection (blue line) of two $T_2$ spheres on two different medial sheets. Middle: A new feature sphere of type $T_3$ is inserted to preserve the internal feature. Right: The internal feature (red line) preserved after our refinement strategy.}
    \label{fig:mm_refine}
\end{figure}

The initialized and updated spheres in the first two steps cannot guarantee there are sufficient spheres sampled on the internal features, such as seams and junctions.
This deficiency usually happens in local regions where an ill-posed connection is caused by two medial spheres that lie on two different medial sheets (see Fig.~\ref{fig:mm_refine}). We first detect these connections under the help of their RPCs, then sample new internal feature spheres using the tangential surface contact points aggregated from two ill-connected spheres. Sec.~\ref{sec:intf_add} explains this refinement in detail. 

\begin{figure}[h]
    \centering
    \includegraphics[width=0.9\linewidth]{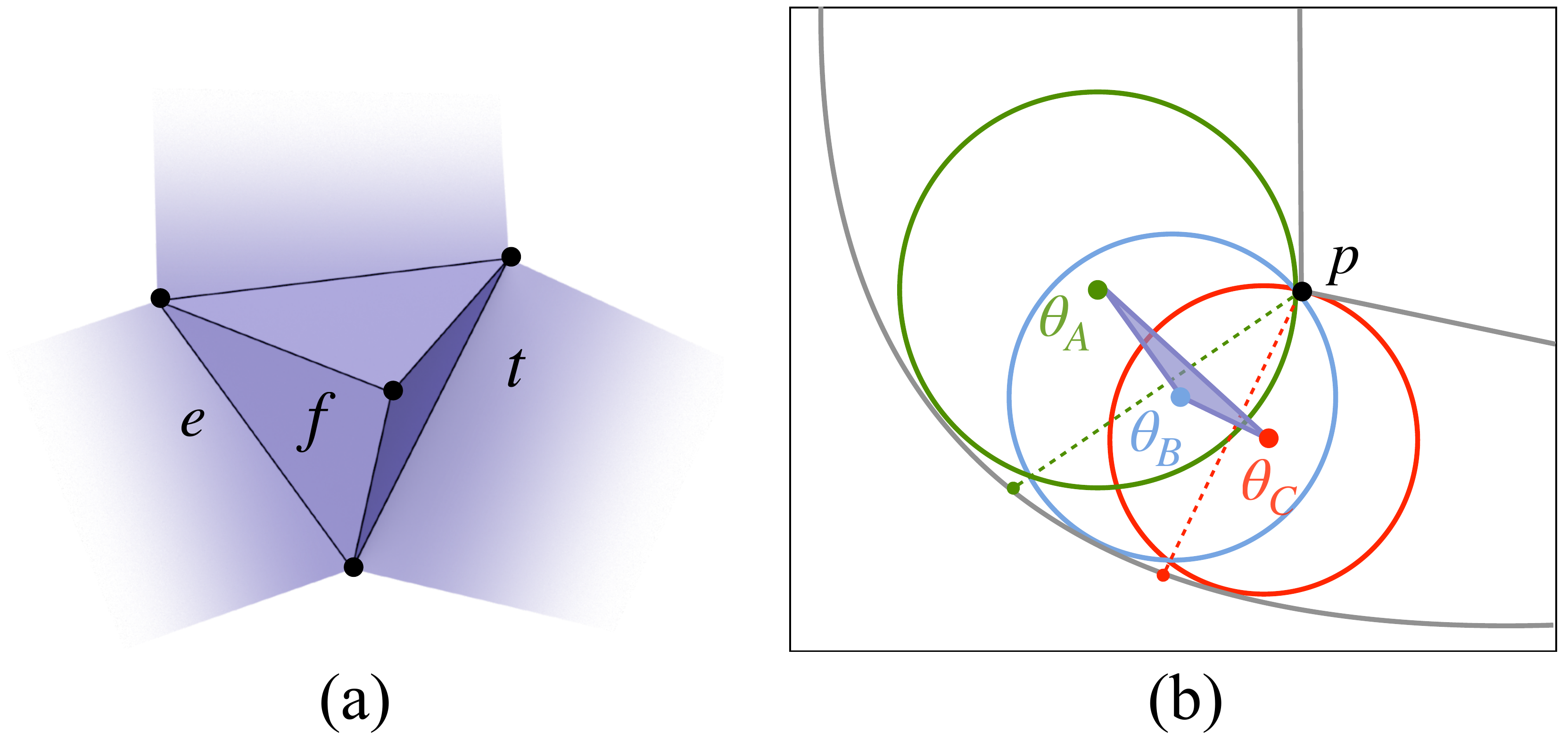}
    \caption{Illustration of some tetrahedron in 3D medial mesh as a 3-dimensional cell of (a), and in 2D as a 2-dimensional cell of (b).}
    \label{fig:refine_thinning_tet}
\end{figure}

The initial medial mesh still contains some flat but solid tetrahedrons as mentioned above. We adapt the thinning algorithm proposed by Liu et al.~\shortcite{liu2010simple} that prunes \textit{simple pairs} of simplices in the medial mesh. A \emph{simple pair} ($x$, $y$) is a pair of simplices such that $y$ is on the boundary of $x$ and there is no other cell in the complex with $y$ on its boundary. Fig.~\ref{fig:refine_thinning_tet} (a) shows a single tetrahedron $t$ in medial mesh as an example, where the tet-face pair ($t$, $f$) is a simple pair, but the face-edge pair ($f$, $e$) is not. It is shown by Ju et al.~\shortcite{ju2007editing} that removing a simple pair does not change the topology even when multiple simple pairs are removed together. This thinning is a pure topological operation that can help us remove tetrahedra from the medial mesh. However, to remove a simple tet-face pair, there could be multiple potential choices. The different order of such removal operation will result in different geometry of the medial mesh, even though their topologies are equivalent. When deciding which simple pair to remove given multiple choices, we shall introduce a quantitative measure for ranking those tetrahedral faces, so that the faces that are less important are prioritized over others to be removed first.

\begin{figure}[h]
    \centering
    \includegraphics[width=0.8\linewidth]{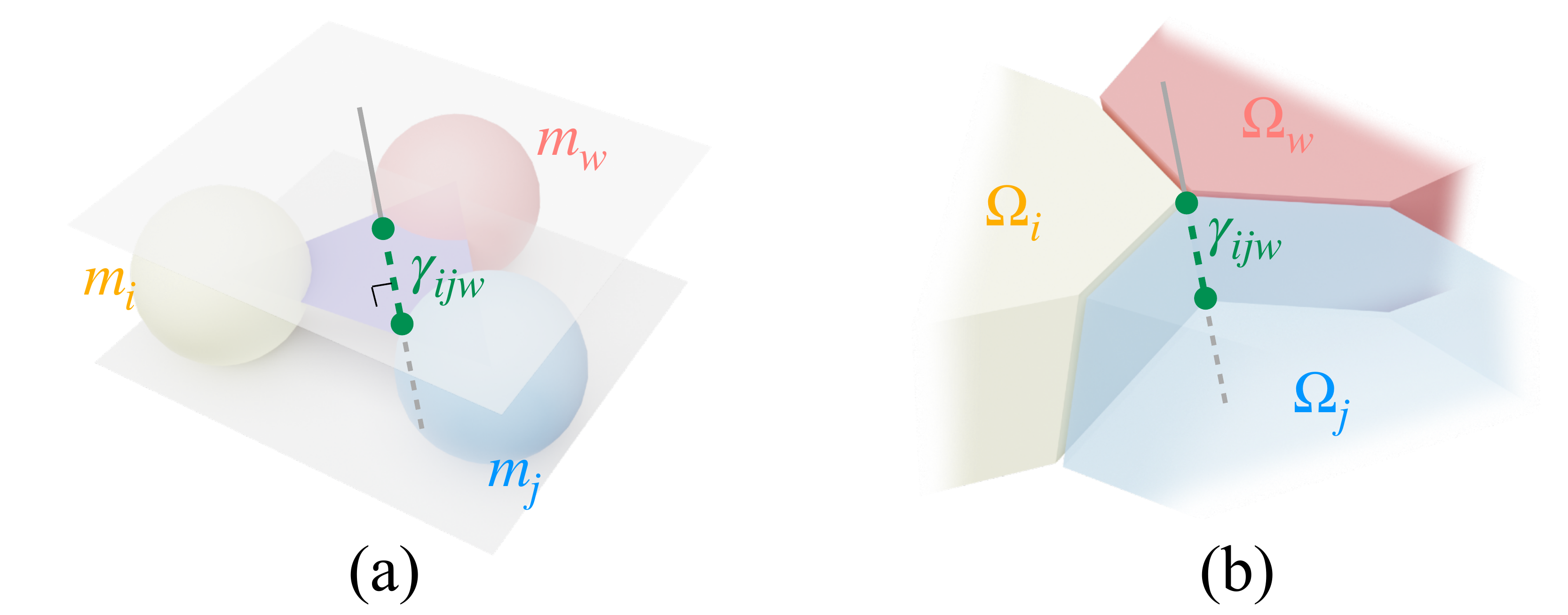}
    \caption{Illustration of \textit{restricted power segment} (RPS). (a) Given a medial face of three medial spheres $\msphere_i$, $\msphere_j$ and $\msphere_w$, and (b) their \textit{restricted power cells} (RPCs), its RPS is shown as the green dotted segment $\rps_{ijw}$.}
    \label{fig:refine_thinning_rps}
\end{figure}

Consider a medial face $f_{ijw}$ connecting three medial spheres $\msphere_i$, $\msphere_j$ and $\msphere_w$ (see Fig.~\ref{fig:refine_thinning_rps}). Since $f_{ijw}$ is a triangle of RT, its dual $\powerseg_{ijw}$ is a line segment of PD. We define the intersection of $\powerseg_{ijw}$ with the shape $\model$ as the \textit{restricted power segment} (RPS) $\rps_{ijw}$:
\begin{equation}
    \rps_{ijw} = \powerseg_{ijw} \cap \model.
\end{equation}
It is worth mentioning that the endpoints of an RPS can be either on the surface $\bmodel$ or inside $\model$. This is because its endpoint could be a dual vertex of an RT tetrahedron, which could be potentially  inside $\model$ (as mentioned in Sec.~\ref{sec:init_mm}).

For medial face $f_{ijw}$, its dual RPS $\rps_{ijw}$ must be perpendicular to the triangle face created by three medial sphere centers ($\mcenter_i$, $\mcenter_j$, $\mcenter_w$). For a local region of medial mesh that is ``thin'', the length of $\rps_{ijw}$ approximates the local thickness of shape, see Fig.~\ref{fig:refine_thinning_rps} the segment in green. Therefore, we defines the importance factor $\alpha_{ijw}$ of a given medial face $f_{ijw}$ as the ratio of the length of $\rps_{ijw}$ over the average diameter of three medial spheres.

For a local medial mesh region that has a flat tetrahedron, every triangle $f_{ijw}$ of the tetrahedron has a dual RPS. We rank these medial triangles by their importance factors $\alpha_{ijw}$ in ascending order. Fig.~\ref{fig:refine_thinning_tet} (b) shows a 2D example where there is a medial triangle (instead of a tetrahedron in 3D) in the neighborhood of a concave sharp feature. The 2D medial triangle ($\msphere_A$, $\msphere_B$, $\msphere_C$) exists when the undesirable edge ($\mcenter_A$, $\mcenter_C$) exists, which means its dual RPS $\rps_{AC}$ exists. In fact, $\rps_{AC}$ degenerates to a point $p$ near the concave sharp feature, making its importance factor $\alpha_{AC}$ to be close to zero. The reason for this degeneracy is because all three spheres pass through the concave sharp feature, making the two RPC of spheres $\msphere_A$ and $\msphere_C$ to be adjacent at the concave point $p$. The other two RPS $\rps_{AB}$ and $\rps_{BC}$ are shown as green and red dotted segments, respectively. Their importance factors $\alpha_{AB}$ and $\alpha_{BC}$ are both close to one. Apparently medial edges $\mcenter_A \mcenter_B$ and $\mcenter_B \mcenter_C$ are more important than edge $\mcenter_A \mcenter_C$ after ranking their dual RPS with their importance factors.

Our \emph{geometry-guided thinning} algorithm starts from raking all  triangles of any tetrahedron in the medial mesh using their importance factors. Then we remove tet-face simple pairs with least importance in each iteration until all tetrahedra are pruned. After removing all tet-face simple pairs, we continue pruning those face-edge simple pairs that belong to the original tetrahedra. To avoid over-pruning 
for models whose medial mesh boundaries are not external features, we use a target important factor $\sigma$ as a stop sign so that the face-edge pair will be not deleted when the importance factor of the current face is beyond this target. Note that $\sigma$ will only impact the pruning of face-edge pairs but not tet-face pairs, the choice of $\sigma$ is discussed in Sec.~\ref{sec:abl_study}.
This pruning operation results in a 3D medial mesh that is ``thin'' with no three dimensional cells, and at the same time, maintains high-quality geometry of the medial mesh. We provide the detailed algorithm as 
Alg.~2 in Supplementary Material, and an ablation study result in Sec.~\ref{sec:abl_study}.

\section{Feature Preservation}
\label{sec:feature_preservation}

The medial feature spheres sampled during initialization (Sec.~\ref{sec:initial}) are not guaranteed to preserve external features after the update of non-feature medial spheres (Sec.~\ref{sec:update}). Also the initial internal features traced in the initial medial mesh (Sec.~\ref{sec:init_mm}) require further refinement (Sec.~\ref{sec:refine}). The connection of two neighboring feature spheres are likely to be destroyed by some non-feature spheres nearby, which results in a fracture of medial features, either external or internal. This is caused mainly by one reason: the feature spheres are not sampled sufficiently in a local region so that the RPC of non-feature spheres may ``invade'' the neighboring RPCs of two feature spheres. Our solution for this issue is to detect the local regions which lack feature spheres, and then add new spheres on them. We will discuss the details of our preservation strategy for both external and internal features.

\subsection{Preserving External Edge Features}
\label{sec:extf_add_se}

\begin{figure}[h]
	\includegraphics[width=\linewidth]{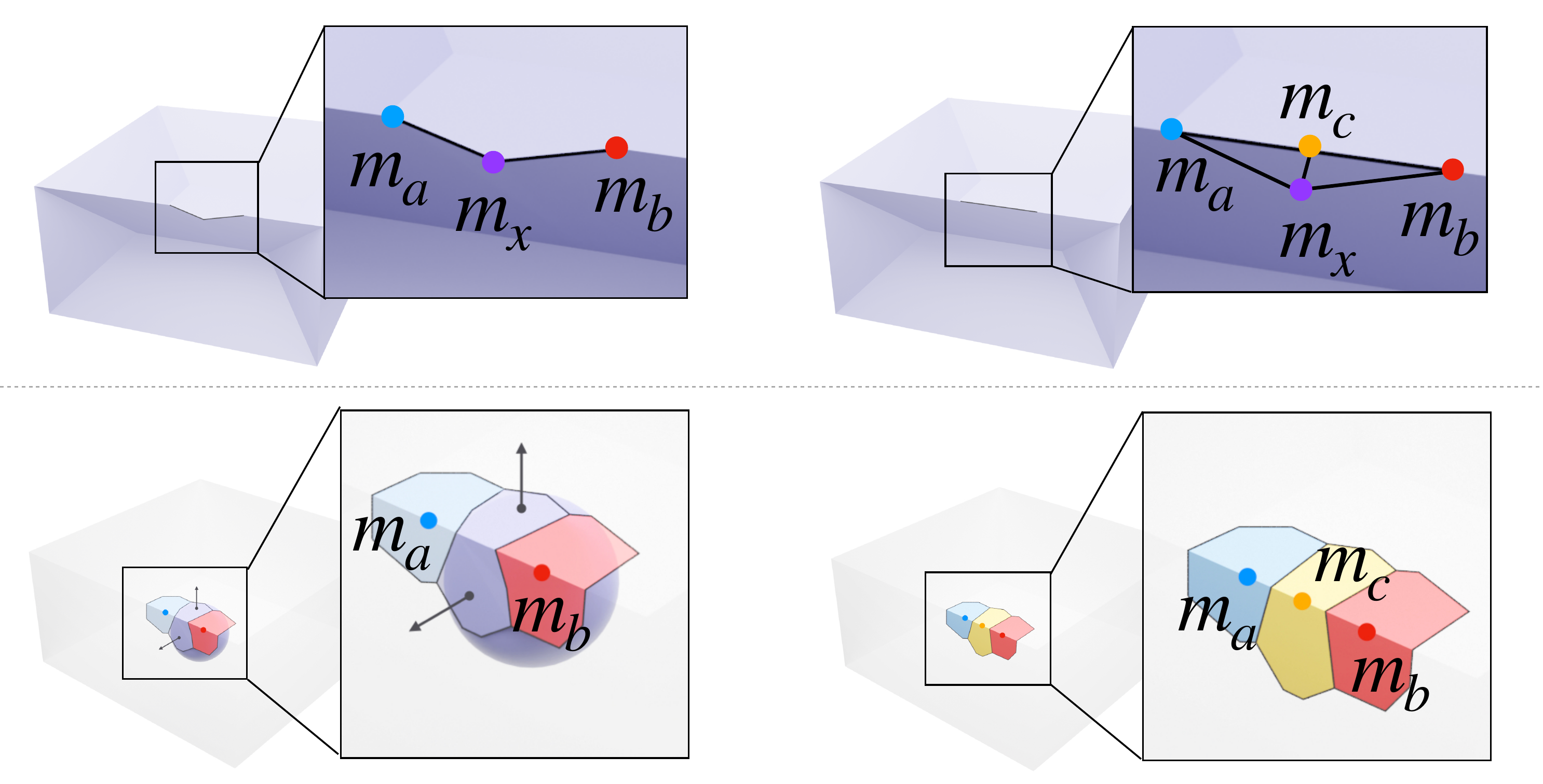}
	\caption{Illustration of how the RPC of a non-feature sphere $\msphere_x$ ``invades'' the RPCs of two neighboring feature spheres $\msphere_a$ and $\msphere_b$ on a sharp edge (bottom left), which could destroy the connectivity of $\msphere_a\msphere_b$ in the medial mesh (top left). We should add a new feature sphere $\msphere_c$ (right) and then recursively check if the sampling is enough.}
	\label{fig:extf_add_se}
\end{figure}

During the generation of initial medial mesh, we sample zero-radius spheres on convex sharp edges to avoid handling infinite sampling density on the input surface. The local external feature could be destroyed if an RPC of any non-feature sphere ``invades'' the RPCs of two neighboring feature spheres on sharp edge (see Fig.~\ref{fig:extf_add_se} left). This is due to the deficiency of feature spheres on external edge features. Therefore, our adaptive re-sampling strategy for maintaining the external features of convex sharp edges is to detect such cases of non-feature spheres then recursively adding new feature spheres until all external edge features are preserved.

We observe that the relationship between RPCs of three spheres can help us find those non-feature spheres that destroy the connectivity of two neighboring zero-radius medial spheres on convex sharp edges. See Fig.~\ref{fig:extf_add_se} for illustrations. For any non-feature medial sphere $\msphere_x = (\mcenter_x, r_x)$ whose RPC is neighboring to the RPCs of two zero-radius medial spheres $\msphere_a=(\mcenter_a,0)$ and $\msphere_b=(\mcenter_b,0)$, where $\mcenter_a\mcenter_b$ is a feature edge to be preserved. The medial sphere $\msphere_x$ breaks the connectivity of $\mcenter_a\mcenter_b$ if the following inequation is true:
\begin{equation}
    \mcenter_x^{\top}\mcenter_x - (\mcenter_a+\mcenter_b)^{\top}\mcenter_x + \mcenter_a^{\top}\mcenter_b \leq r_x^2.
\end{equation}
The detailed proof is given in the Supplementary Material Sec.~2.

\begin{wrapfigure}{r}{0.1\textwidth}
  \centering
  \vspace{-5pt}
  \begin{center}
    \includegraphics[width=0.1\textwidth]{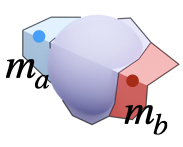}
  \end{center}
  \vspace{-5pt}
\end{wrapfigure}
For every non-feature sphere $\msphere_x$, we search for its $\numneighbors$-nearest external edge features $\msphere_{a_k}\msphere_{b_k}$ ($k=1...\numneighbors$) and check if $\msphere_x$ destroy the connectivity of them. 
If it breaks, we add a new feature sphere $\msphere_{c_k}$ in the middle of edge $\msphere_{a_k}\msphere_{b_k}$ and recursively check edges $\msphere_{a_k}\msphere_{c_k}$ and $\msphere_{b_k}\msphere_{c_k}$. This recursion will be infinite only when $\msphere_x$ protrudes from the convex sharp edge (see the right inset figure) and any new feature sphere $\msphere_{c_k}$ could be contained inside $\msphere_x$. Therefore in this case the medial sphere $\msphere_x$ should be removed. However, this is unlikely to happen since all non-feature spheres have been updated to their ideal position in the second step, so they would not protrude from the convex sharp edges. 

\subsection{Preserving External Corner Features}
\label{sec:extf_add_corner}

\begin{figure}[h]
    \centering
    \includegraphics[width=\linewidth]{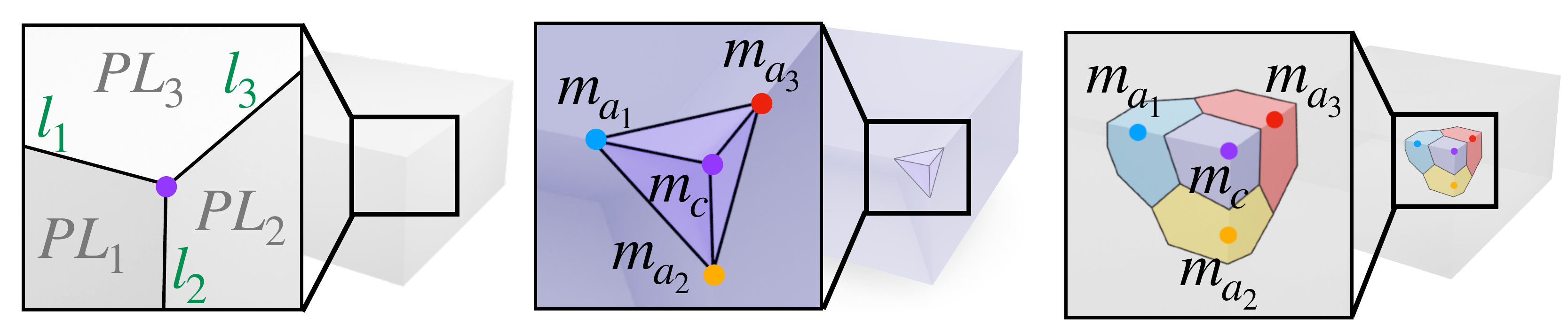}
    \caption{Left: A convex corner formed by three convex edges $\ccl_i$ and three incident planes $\tanpl_i$ ($i$=1,2,3). Middle: The structure of the ``corner cap'' where any two neighboring external feature spheres $\msphere_{a_{i}}$ and $\msphere_{a_{j}}$ are connected, created a medial face of $\bigtriangleup (\msphere_c \msphere_{a_{i}} \msphere_{a_{j}}$). Right: The surface RPC of these four external feature spheres.}
    \label{fig:extf_add_corner_problem}
\end{figure}

\begin{figure}[h]
    \centering
    \includegraphics[width=\linewidth]{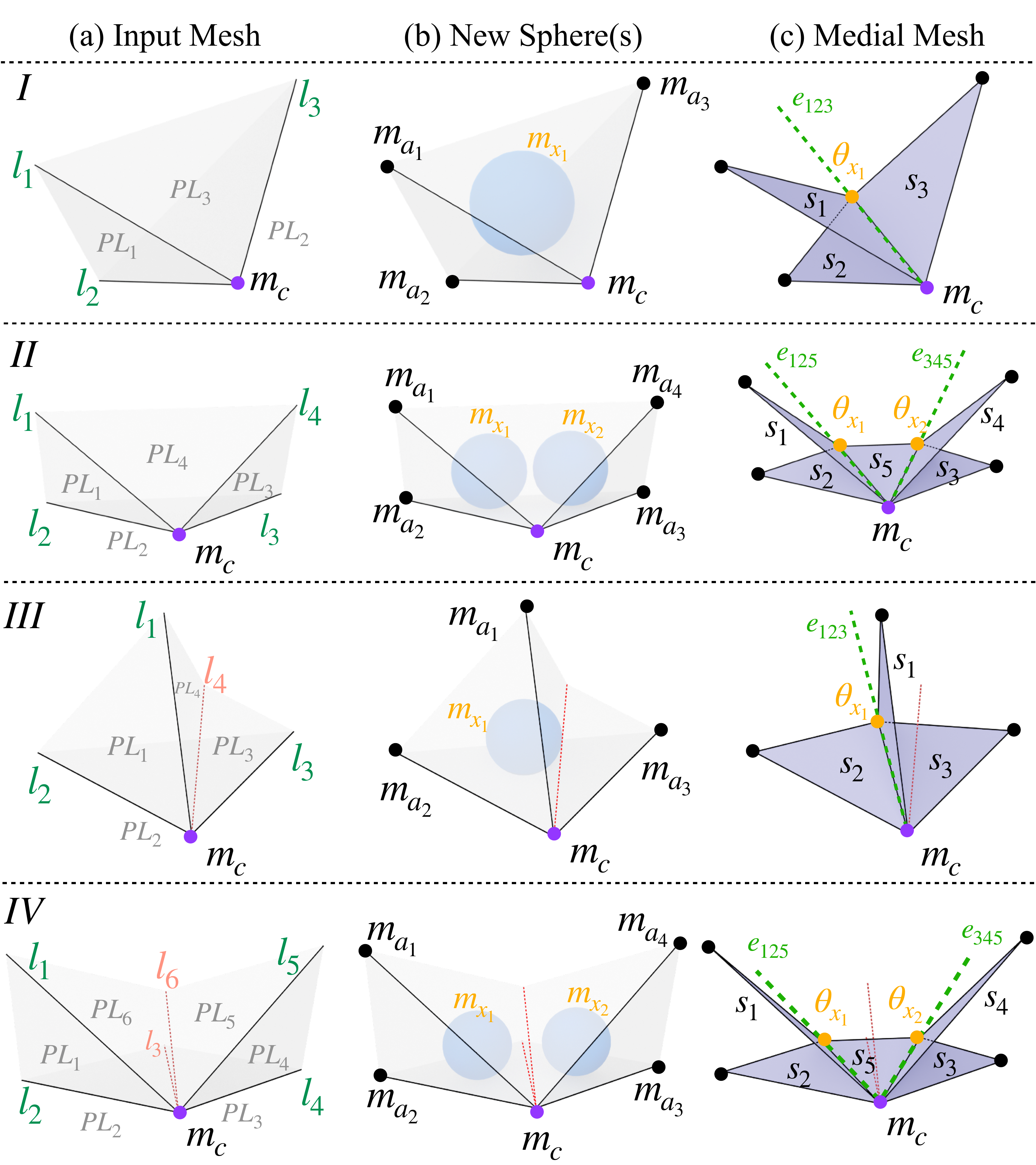}
    \vspace{-10pt}
    \caption{Some examples of corners incident to convex and/or concave sharp edges. For each example, we show their valid seams represented as tangent collections symbolically. Row $\Romannum{1}$: $e_{123}$ = $(\tanpl_1, \tanpl_2, \tanpl_3)$. Row $\Romannum{2}$: (1) $e_{125}$ = $(\tanpl_4, \tanpl_1, \tanpl_2)$; (2) $e_{345}$ = $(\tanpl_2, \tanpl_3, \tanpl_4)$. Row $\Romannum{3}$: $e_{123}$ = $(\tanpl_2, \ccl_4, \tanpl_1)$. Row $\Romannum{4}$: (1) $e_{125}$ = $(\tanpl_6, \tanpl_1, \tanpl_2)$; (2)  $e_{345}$ = $(\tanpl_3, \tanpl_4, \tanpl_5)$.}
    \label{fig:extf_add_corner}
    \vspace{-10pt}
\end{figure}

An external corner is typically formed by three or more sharp edges, including either convex edges or concave edges. As shown in Fig.~\ref{fig:extf_add_corner_problem}, zero-radius medial spheres sampled on one sharp convex edge may get entangled with the ones on neighboring sharp convex edges, which may result in a ``corner cap'' in the resulting medial mesh. Since each sharp edge $\ccl_k$ is formed by two adjacent planes $\tanpl_{k_1}$ and $\tanpl_{k_2}$, it must define a sheet of medial axis passing through this sharp edge, with medial spheres tangent to its two adjacent planes. Three sheets potentially intersect and form a seam. For example in Fig.~\ref{fig:extf_add_corner} $\Romannum{1}$, any convex edge $\ccl_k$ forms a sheet $s_k$ on medial axis, and three sheets join at a seam $e_{123}$. 

One possible solution to remove the ``cap'' around corners is to directly sample new medial spheres on potential seams using our internal feature preservation strategy described in Sec.~\ref{sec:intf_add}. However, such addition of new spheres on seams might break the connectivity of sharp edges nearby (Sec.~\ref{sec:extf_add_se}), which would require adding more samples on sharp edges in order to preserve them. This would cause an infinite loop and the sphere sampling density could run into infinity, as two sharp edges are approaching closer to a corner. To solve this challenge, we select a small region within distance $\delta$ from the given corner $\msphere_c$, and propose a corner preservation strategy to approximate medial mesh structures in this small region around the corner. Our corner preservation scheme works by recursively tracing sheets of medial axis, starting from those convex sharp edges, until we found their intersecting seams. It consists of the following three steps.


\subsubsection{Enumerating Initial Set of Sheets.}
We first sort neighboring sharp edges $\ccl_k$ ($k=1...N$), both convex and concave, together with their adjacent planes $\tanpl_k$ in a counter-clockwise order. In this way we can enumerate an initial set of potential sheets using the tangent planes incident to all convex sharp edges. Note that concave sharp edges will not contribute any new sheet as they only create smooth transitioning conic sections on existing sheets, and spheres tangential to only one concave edge are of type $T_1^c$ as spikes.

Take Fig.~\ref{fig:extf_add_corner} $\Romannum{3}$ as an example. Corner $\msphere_c$ is where three convex edges ($\ccl_1$, $\ccl_2$ and $\ccl_3$ in green) and one concave edge ($\ccl_4$ in red) converge. Also $\msphere_c$ has four incident planes $\tanpl_k$, $k \in \{1...4\}$, where $\tanpl_3$ and $\tanpl_4$ are shared by a concave edge $\ccl_4$. There are three possible sheets $s_k$ traced from convex edges $\ccl_k$, $k=1,2,3$. For example, sheet $s_1$ is formed by a set of medial spheres tangential to two tangent planes $\tanpl_4$ and $\tanpl_1$ adjacent to convex edge $\ccl_1$. So we can have symbolic representations for these sheets: $s_1$ = ($\tanpl_4$, $\tanpl_1$), $s_2$ = ($\tanpl_1$, $\tanpl_2$), and $s_3$ = ($\tanpl_2$, $\tanpl_3$). 

\subsubsection{Finding Seams and New Sheets Recursively.}
If two neighboring sheets in counter-clockwise order intersect, we will be able to find a medial sphere on the intersecting seam through their aggregated set of tangent planes and/or concave edges. In addition, we can find out another new sheet that intersects with them on the same seam. In the example of Fig.~\ref{fig:extf_add_corner} $\Romannum{3}$, we can find their potential intersecting seams as follows:
\begin{enumerate}
    \item Checking potential intersection between $s_1$ and $s_2$:
    \begin{enumerate}
        \item[(a)] Using tangent collection ($\tanpl_4$, $\tanpl_1$, $\tanpl_2$) or
        \item[(b)] Using tangent collection ($\ccl_4$, $\tanpl_1$, $\tanpl_2$);
    \end{enumerate}
    \item Checking potential intersection between $s_3$ and $s_1$: 
    \begin{enumerate}
        \item[(c)] Using tangent collection ($\tanpl_1$, $\tanpl_2$, $\tanpl_3$) or
        \item[(d)] Using tangent collection ($\tanpl_1$, $\tanpl_2$, $\ccl_4$).
    \end{enumerate}
\end{enumerate}
Here (b) and (d) defines the same seam symbolically so we can merge them as $e_{123}$. 
Each tangent collection of three elements (tangent planes and/or concave edges) imply a potential medial sphere on the seam of medial axis. We can obtain the sphere $\msphere_{x_i}$ by minimizing the energy of Eq.~\eqref{eq:energy_updated} as described in Sec.~\ref{sec:update}. If the optimized energy is larger than threshold $\epsilon$, it means there does not exist a sphere that can be tangential to these tangent collections. The example in $\Romannum{3}$ only have one valid seam sphere based on collections (b) and (d), while the other two collections (a) and (c) cannot lead to any medial sphere based on their optimized energies. Once a seam sphere is computed on collection (b), based on the tangent information we can obtain a new sheet ($\ccl_4$, $\tanpl_2$) that is intersecting with both $s_1$ and $s_2$ on the same seam. In this case the new sheet is exactly $s_3$=($\tanpl_2$, $\tanpl_3$) because the concave edge $\ccl_4$ and tangent plane $\tanpl_3$ are adjacent, so it will not be included further. 

This process is implemented as a circular queue of potential sheets ordered counter-clockwise. Every time we pop out a sheet, we check its intersection with the next sheet in the queue. If two of them forms a seam, we sample a new medial sphere with two sheets' info recorded, and push the new sheet candidate (if unique) into the queue. Otherwise, we push it back into the queue and pop out the next sheet to check. This process continues until the queue is empty.


\subsubsection{Connect the Spheres on Sheets and Seams to form a Medial Mesh around the Corner.}
Once all seams are found with their medial spheres sampled, our third step is to construct an approximated medial mesh around the corner. 
For each convex sharp edge, we sample a zero-radius medial sphere at distance $\delta$ from the corner. For example, in Fig.~\ref{fig:extf_add_corner} $\Romannum{4}$ we sample $\msphere_{a_1}$, $\msphere_{a_2}$, $\msphere_{a_3}$, and $\msphere_{a_4}$ on convex edges $\ccl_1$, $\ccl_2$, $\ccl_4$, and $\ccl_5$, respectively. For each sheet found in the above process, we connect its two medial spheres with the corner to form a medial triangle, \eg, $\msphere_{x_1}$, $\msphere_{x_2}$ and the corner $\msphere_{c}$ form a triangle for the sheet $s_5$; $\msphere_{x_1}$, $\msphere_{a_1}$ and the corner $\msphere_{c}$ form a triangle for the sheet $s_1$, etc. 

In this way all sheets around the corner can be constructed as medial triangles, with all seams represented as edges between neighboring triangles. In Sec.~\ref{sec:results_more} we show our experiment results on shapes with different corner cases. Note that a corner could be just the tip of a cone shape without any neighboring sharp edges (see Fig.~\ref{fig:saddle} (e)), in which case our medial mesh computed with RRT in Sec.~\ref{sec:init_mm} can already preserve it as long as there is a zero-radius medial sphere sampled on the tip.

\subsection{Preserving Internal Features}
\label{sec:intf_add}

\begin{figure}[h]
	\includegraphics[width=0.9\linewidth]{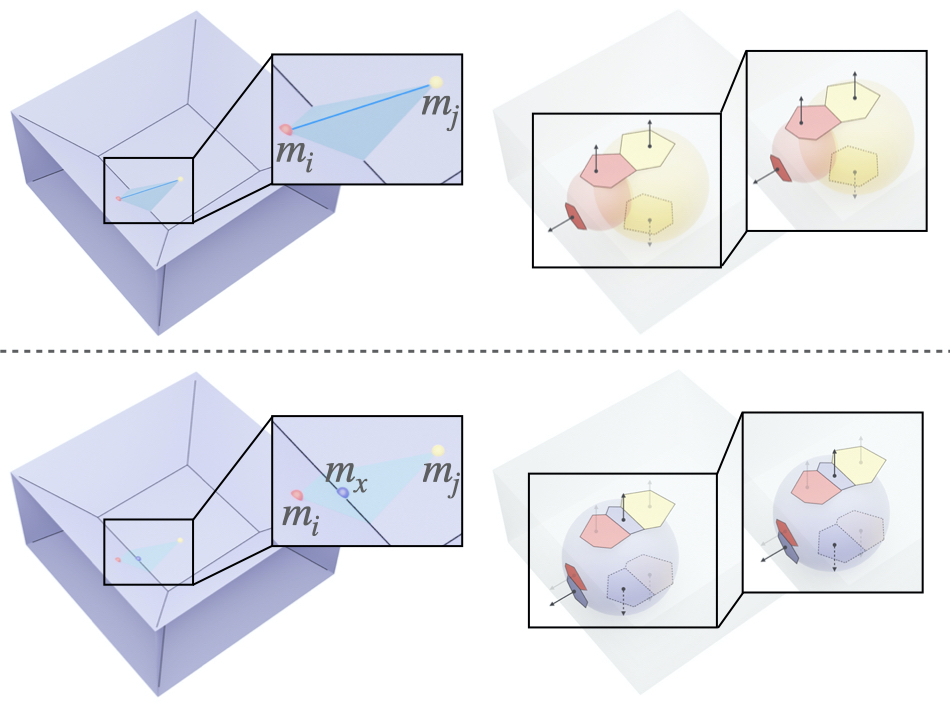}
	\vspace{-5pt}
	\caption{The local region near two medial spheres $\msphere_i$ and $\msphere_j$ that are on two different medial sheets. There is a lack of feature spheres in this local region, causing $\msphere_i$ and $\msphere_j$ to be connected (top). A new feature sphere $\msphere_x$ is added to preserve the internal feature of medial axis (bottom).}
	\label{fig:intf_add}
\end{figure}

Internal feature preservation requires sufficient sampling of internal feature spheres. Therefore, our initial internal features detected from the initial medial mesh need to be refined after detecting the regions that lack feature spheres. We have observed that the deficiency of internal feature spheres normally occurs when two connected non-feature medial spheres belong to different medial sheets. Based on this observation, we propose an internal sphere insertion strategy that is able to directly insert medial spheres without relying on additional surface samples (for computing Voronoi balls). 

Suppose two medial spheres $\msphere_i$ and $\msphere_j$ are connected in the medial mesh. We are expecting every CC of $\msphere_i$ must be adjacent to one corresponding CC of $\msphere_j$, which means sphere $\msphere_i$ and $\msphere_j$ are on the same medial sheet. Otherwise they should not be connected since they are on different medial sheets. A deficiency of feature spheres is detected in this local region if $\msphere_i$ and $\msphere_j$ belong to two different medial sheets on the medial mesh. In this case, a new feature sphere $\msphere_x$ will be inserted using the aggregated connected components of $\msphere_i$ and $\msphere_j$. Fig.~\ref{fig:intf_add} shows an illustration of such case. Two $T_2$ medial spheres $\msphere_i$ and $\msphere_j$ are connected so that the internal feature (black solid line) cannot be preserved as expected. Only one CC of $\msphere_i$ is adjacent to one CC of $\msphere_j$, and their another CCs are on different surface regions and not adjacent to each other (see Fig.~\ref{fig:intf_add} right column). This indicates that $\msphere_i$ and $\msphere_j$ are on two different medial sheets, and there is a deficiency of feature spheres in this local region. Using the aggregated CCs of these two medial spheres, we can insert a new medial sphere $\msphere_x$ of type $T_3$ that has tangential contacts with the surface at three points using the same method in Sec.~\ref{sec:update}.

\section{Experiments}
\label{sec:exp}

We implement our algorithm in C++, using CGAL for triangulation calculation, and Eigen for linear algebra routines. The exact calculation of restricted power diagrams extends the Voronoi package in Geogram\footnote{Geogram: http://alice.loria.fr/software/geogram/doc/html/index.html}. We run our experiments on a computer with a 3.60GHz Intel(R) Core(TM) i7-9700K CPU and 32 GB memory. All models used in this paper are from the ABC dataset \cite{Koch_2019_CVPR} and their sizes are normalized to the [0, 10] range.

\begin{figure}[t]
    \centering
    \includegraphics[width=\linewidth]{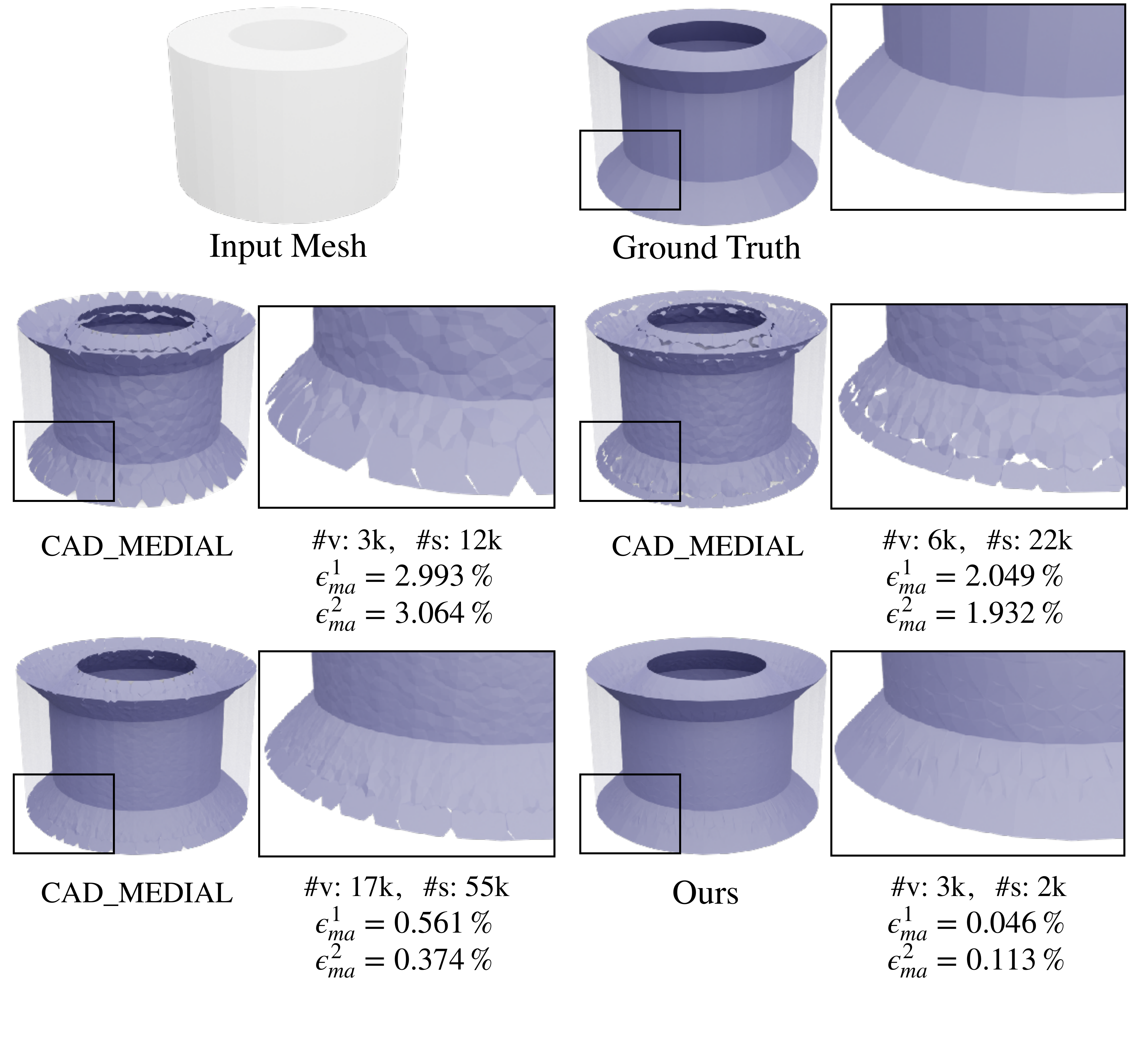}
    \vspace{-30pt}
    \caption{Comparison with CAD\_MEDIAL \cite{dey2003approximate} on a CAD model with ground truth medial axis.}
    \label{fig:comp_medialCAD}
\end{figure}

\paragraph{Evaluation Metrics.} We use the two-sided Hausdorff distance error, denoted as $\epsilon$, to assess the surface reconstruction accuracy using the generated medial meshes. $\epsilon^{1}$ is the one-sided Hausdorff distance from the original surface to the surface reconstructed from MAT, and $\epsilon^{2}$ is the distance in reverse side. 
We also directly evaluate the difference between the approximated medial mesh and the ground-truth medial axis for some input surfaces. The $\epsilon_{ma}^{1}$ is the one-sided Hausdorff distance from the ground-truth medial axis to the approximated medial mesh, and $\epsilon_{ma}^{2}$ is the distance in reverse side. Note that we manually generate the ground-truth medial axis for some simple shapes, such as those in Fig.~\ref{fig:intro_dode}, Fig.~\ref{fig:comp_medialCAD}, and Fig.~\ref{fig:comp_gt}.
For sampling-based methods, we also show $\#v$ as the number of surface samples used. We show $\#s$ as the number of medial spheres for the medial meshes generated from each method. All Hausdorff distances are evaluated as percentages of the distances over the diagonal lengths of the models' bounding boxes.

\begin{figure*}[h]
    \centering
    \includegraphics[width=0.95\linewidth]{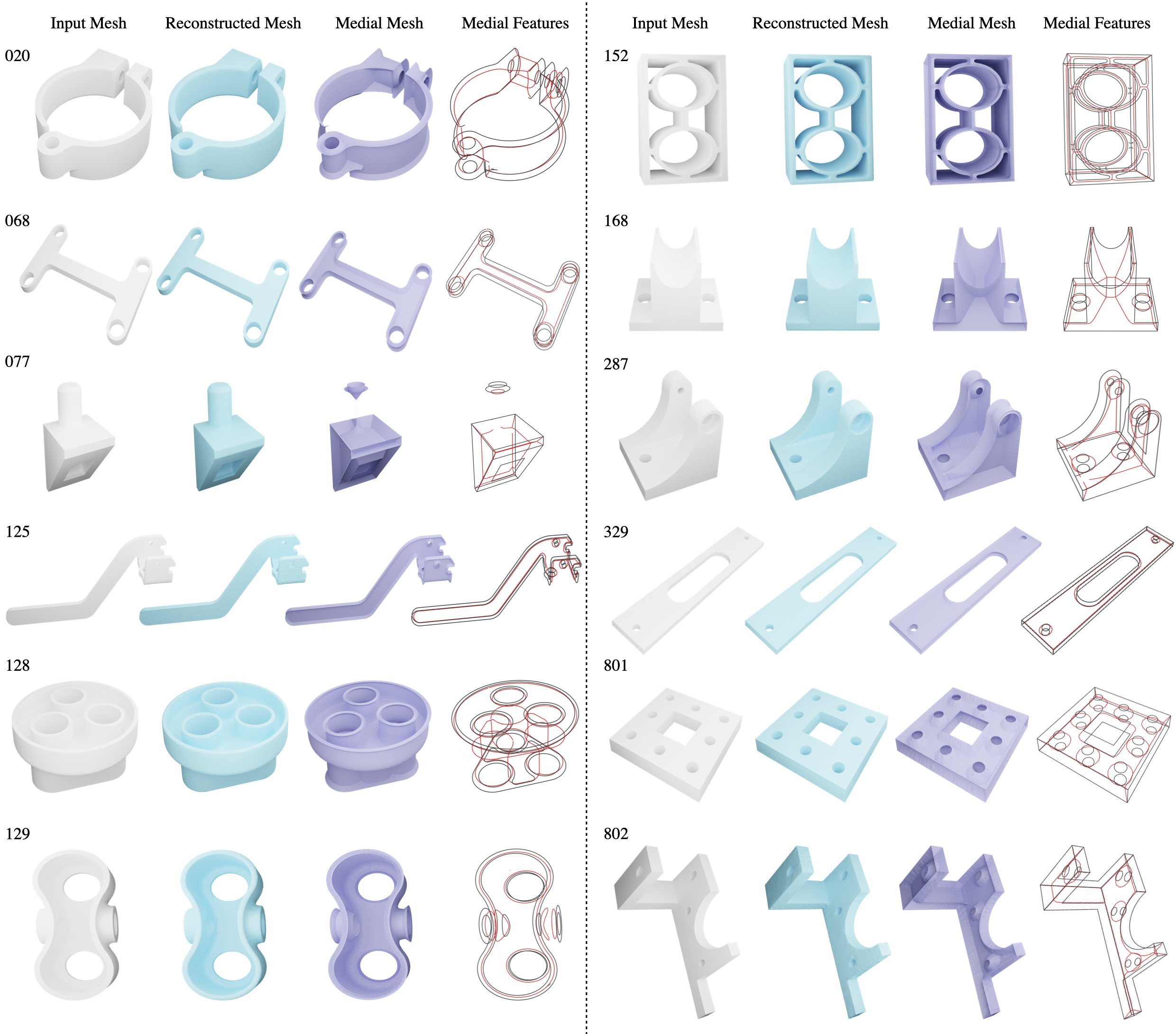}
    \vspace{-5pt}
    \caption{From left to right are the input surface meshes, the surfaces reconstructed from our medial meshes, the generated medial meshes, and the extracted medial features. For the medial features, the black curves are the external features and the red curves are the internal features. For reconstructed surfaces, their reconstruction errors are given in Table~\ref{tab:more_rec_error}.}
    \label{fig:more_rec}
\end{figure*}

We show a set of qualitative results on various 3D CAD models in Fig.~\ref{fig:more_rec} with their running time statistics summarized in Table~\ref{tab:time}. More detailed views of those computed medial mesh and their extracted medial features are shown in the supplementary video. The shape reconstruction errors of those models listed in Fig.~\ref{fig:more_rec}, measured by the Hausdorff distance error, are given in Table~\ref{tab:more_rec_error}.

\begin{table}[t!]
\small
\caption{Statistics of our running time in seconds. $\#v^{\ast}$ is the number of vertices of original model, $\#v$ is the number of surface samples used to generate initial Voronoi balls, $\#f$ is the number of triangle faces in the input mesh, $\#s$ is the number of generated medial spheres, and $\#t$ is the number of active tetrahedrons before thinning process in Sec.~\ref{sec:refine}. Note that the running time of calculating surface RPD relates to $\#v$, $\#f$, and $\#s$ using the clipping algorithm \cite{yan2009isotropic}. And the running time of thinning process in S4 relates to $\#t$. S1 is the running time of calculating initial medial sphere centers (Sec.~\ref{sec:initial}). S2 is the running time of updating medial spheres (Sec.~\ref{sec:update}). S3 is the running time of calculating initial medial mesh (Sec.~\ref{sec:init_mm}). S4 the running times for refining medial mesh (Sec.~\ref{sec:refine}). The model's ID\# corresponds to those shown in Fig.~\ref{fig:teaser} and Fig.~\ref{fig:more_rec}.}
\vspace{-10 pt}
\begin{center}
\scalebox{0.9}{
\begin{tabular}{|c|c|c|c|c|c|c|c|c|c|c|}
\whline{0.75pt}
Model & $\#v^{\ast}$ & $\#v$ & $\#f$ & $\#s$ & $\#t$ & S1 & S2 & S3 & S4 & Total \\
\whline{0.75pt}
020 & 21k & 5k & 41k & 11k & 4k  & 2.8 & 1.4 & 4.9  & 6.8   & 15.9 \\
068 & 7k  & 2k & 14k & 5k  & 773 & 0.8 & 0.3 & 1.2  & 4.6   & 8.9 \\
077 & 19k & 5k & 38k & 11k & 17k & 2.5 & 2.1 & 4.7  & 50.3  & 59.6 \\
125 & 5k  & 5k & 10k & 6k  & 4k  & 0.7 & 0.5 & 1.1  & 7.6   & 9.9 \\
128 & 25k & 5k & 51k & 31  & 28k & 3.8 & 9.8 & 12.6 & 84.5  & 110.7 \\
129 & 21k & 4k & 42k & 18k & 11k & 2.8 & 2.6 & 6.4  & 16.5  & 23.1 \\
152 & 41k & 9k & 82k & 39k & 37k & 6.3 & 4.8 & 27.1 & 118.6 & 156.8 \\
168 & 22k & 4k & 44k & 11k & 9k  & 2.8 & 1.0 & 5.5  & 14.5  & 23.8 \\
287 & 22k & 5k & 45k & 12k & 12k & 2.9 & 1.7 & 6.5  & 23.2  & 34.3 \\
329 & 5k  & 2k & 9k  & 6k  & 507 & 0.7 & 0.3 & 1.5  & 4.2   & 6.7 \\
801 & 17k & 4k & 34k & 9k  & 4k  & 2.0 & 0.9 & 4.3  & 6.3   & 13.5 \\
802 & 11k & 3k & 21k & 6k  & 5k  & 1.3 & 0.7 & 2.5  & 6.7   & 11.2 \\
\whline{0.75pt}
\end{tabular}}
\end{center}
\label{tab:time}
\vspace{-5 pt}
\end{table}

\begin{table}[h]
\small
\caption{Quantitative comparison on shape reconstruction error among PC \cite{amenta2001power}, SAT \cite{miklos2010discrete}, VC \cite{yan2018voxel} and ours. $\#s$ is the number of generated medial spheres, $\epsilon$ is the two-sided Hausdorff distance between original surface and reconstruction (maximum of $\epsilon^1$ and $\epsilon^2$ described in Sec.~\ref{sec:exp}), and $E$ is the Euler characteristic. We also show ground truth Euler characteristic as ``GT $E$''. For CAD models, our method always gives the best reconstruction results with correct Euler characteristics, and with smaller amount of medial spheres generated.}
\vspace{-10 pt}
\begin{center}
\scalebox{0.75}{
\begin{tabular}{|c|c|c|c|c|c|c|c|c|c|c|c|c|}
\hline
Model & \multicolumn{3}{c|}{PC} & \multicolumn{3}{c|}{SAT} & \multicolumn{3}{c|}{VC} & \multicolumn{3}{c|}{Ours} \\
(GT $E$) & $\#s$ & $\epsilon$ & $E$ & $\#s$ & $\epsilon$ & $E$ & $\#s$ & $\epsilon$ & $E$ & $\#s$ & $\epsilon$ & $E$ \\
\hline
020	(-1)  &   35k &   0.733 & 78k & 137k  &   0.715 & 97  & 34k   & 1.319 &	\textbf{-1}  & 11k   & \textbf{0.400} & \textbf{-1} \\ \hline
068	(-3)  &   11k &   0.577 & 20k & 85k   &   0.572 & 44  & 77k   & 0.722 &	\textbf{-3}  & 5k    & \textbf{0.550} & \textbf{-3} \\ \hline
077	( 1)  &   32k &   2.508 & 86k & 235k  &   1.492 & 140 & 72k   & 2.822 &	\textbf{ 1}  & 11k   & \textbf{1.381} & \textbf{ 1} \\ \hline
125	(-2)  &   7k  &   0.708 & 8k  & 51k   &   0.406 & 31  & 10k   & 0.501 &	\textbf{-2}  & 6k    & \textbf{0.198} & \textbf{-2} \\ \hline
128	(-2)  &   33k &   0.709 & 61k & 294k  &   0.721 & 355 & 126k  & 0.845 &	\textbf{-2}  & 31k   & \textbf{0.695} & \textbf{-2} \\ \hline
129	(-3)  &   31k &   0.866 & 58k & 231k  &   0.676 & 192 & 56k   & 1.033 &	\textbf{-3}  & 18k   & \textbf{0.640} & \textbf{-3} \\ \hline
152	(-7)  &   53k &   1.490 & 96k & 455k  &   1.476 & 364 & 111k  & 1.220 &	\textbf{-7}  & 39k   & \textbf{0.787} & \textbf{-7} \\ \hline
168	(-1)  &   39k &   1.383 & 98k & 263k  &   0.810 & 235 & 86k   & 1.716 &	\textbf{-1}  & 11k   & \textbf{0.632} & \textbf{-1} \\ \hline
287	(-3)  &   38k &   1.842 & 93k & 272k  &   0.828 & 113 & 150k  & 7.902 &	\textbf{-3}  & 12k   & \textbf{0.681} & \textbf{-3} \\ \hline
329	(-2)  &   37k &   1.737 & 65k & 57k   &   1.199 & 65  & 29k   & 0.634 &	\textbf{-2}  & 6k    & \textbf{0.468} & \textbf{-2} \\ \hline
801	(-8)  &   27k &   1.113 & 55k & 202k  &   0.706 & 41  & 61k   & 1.591 &	\textbf{-8}  & 9k    & \textbf{0.405} & \textbf{-8} \\ \hline
802	(-3)  &   16k &   1.240 & 39k & 128k  &   0.950 & 24  & 17k   & 2.195 &	\textbf{-3}  & 6k    & \textbf{0.465} & \textbf{-3} \\
\hline
\end{tabular}}
\end{center}
\label{tab:more_rec_error}
\vspace{-5 pt}
\end{table}


\subsection{Comparison with the \emph{CAD\_MEDIAL} Method}


To our best knowledge, CAD\_MEDIAL \cite{dey2003approximate} is the only method that works on preserving external features of medial axis so far. 
However, their sampling condition near sharp edges is extremely strict and very hard to be achieved if the sharp edge is not a straight line. 
Fig.~\ref{fig:comp_medialCAD} shows the visual comparison of the medial mesh quality of ours and CAD\_MEDIAL with different numbers of surface samples used on a CAD model. It also shows the Hausdorff distances w.r.t. its ground truth medial axis. It can be seen that increasing the number of surface samples for CAD\_MEDIAL would make their generated medial mesh more complete, however, the sampling density around sharp edges needs fine tuning and very likely to generate incomplete structures. In contrast, our method does not require strict sampling condition around those non-smooth surface regions and produces more accurate approximation of convex external features with smaller Hausdorff distances.




\begin{figure}[h]
    \centering
    \includegraphics[width=\linewidth]{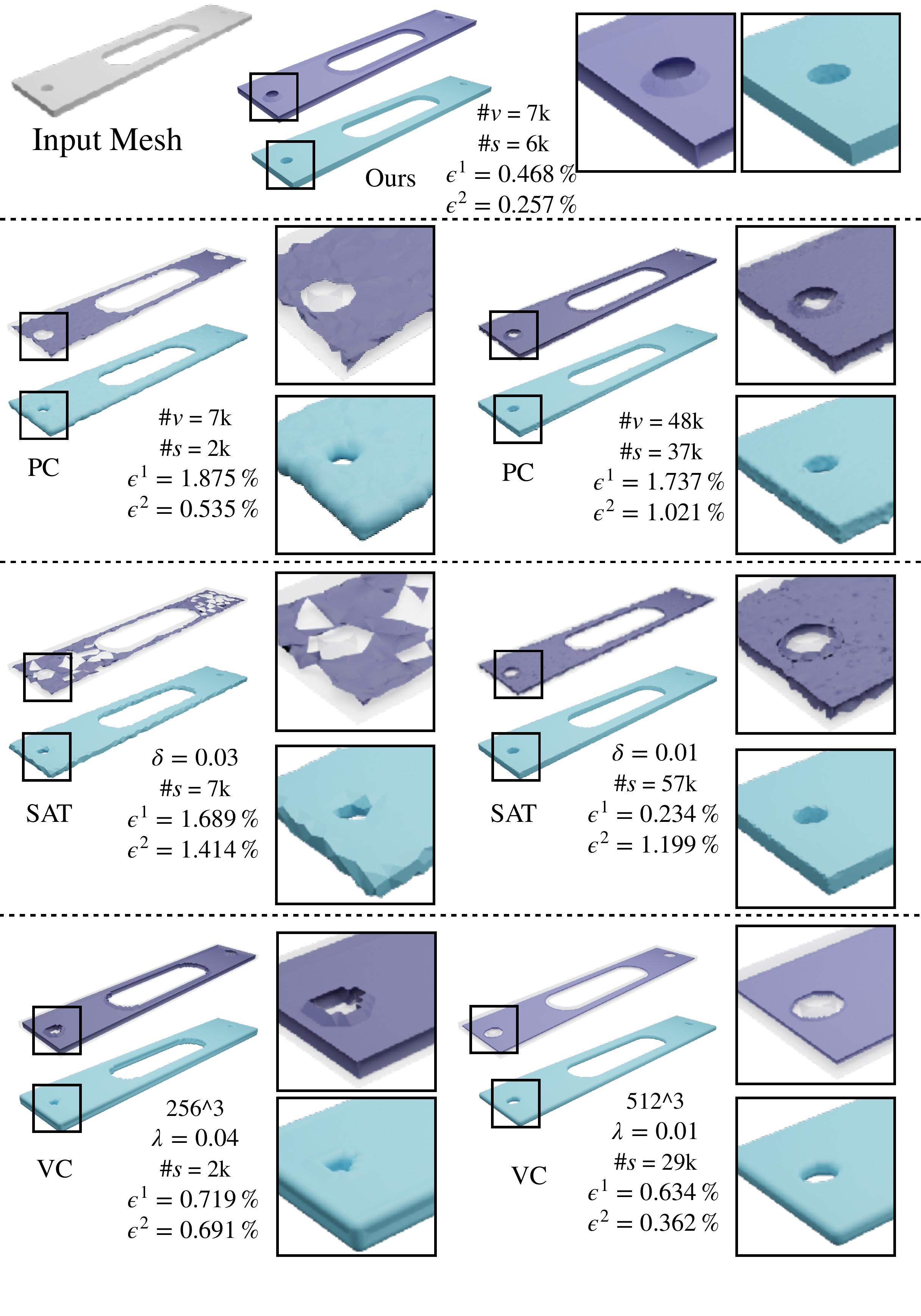}
    \vspace{-5pt}
    \caption{Qualitative comparison of the medial mesh and quantitative comparison of the reconstructed mesh among our method and two sampling-based methods: PC \cite{amenta2001power} and SAT \cite{miklos2010discrete}, and a voxel-based method VC \cite{yan2018voxel}.}
    \label{fig:comp_recon}
\end{figure}

\begin{figure}[h]
    \centering
    \includegraphics[width=\linewidth]{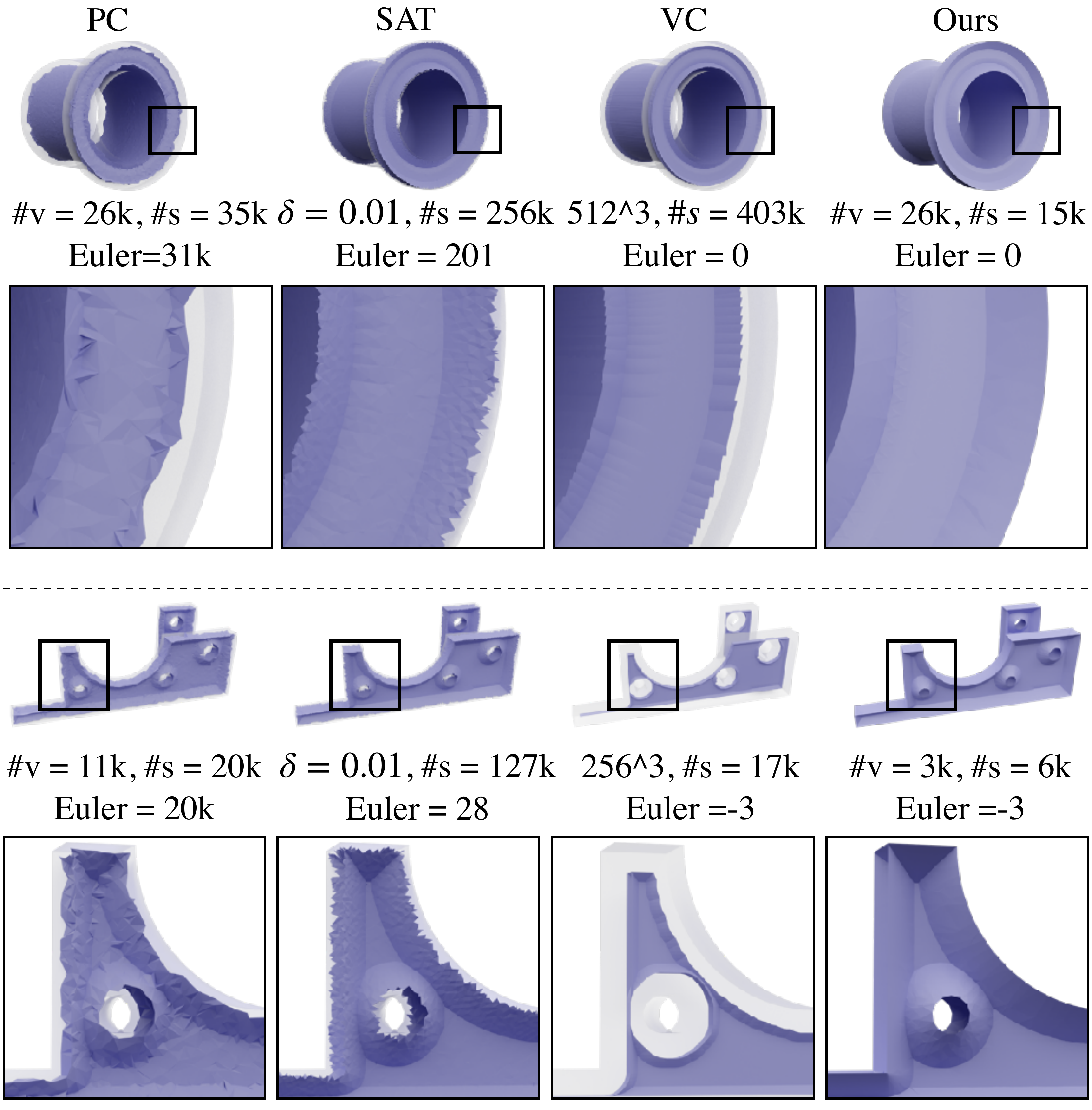}
    \vspace{-5pt}
    \caption{Comparing the feature preserving ability of our method with two sampling-based methods: PC \cite{amenta2001power} and SAT \cite{miklos2010discrete}, also a voxel-based method VC \cite{yan2018voxel}.}
    \label{fig:comp_features}
\end{figure}

\subsection{Comparison with Sampling-based Methods}

\begin{figure}[h]
    \centering
    \includegraphics[width=\linewidth]{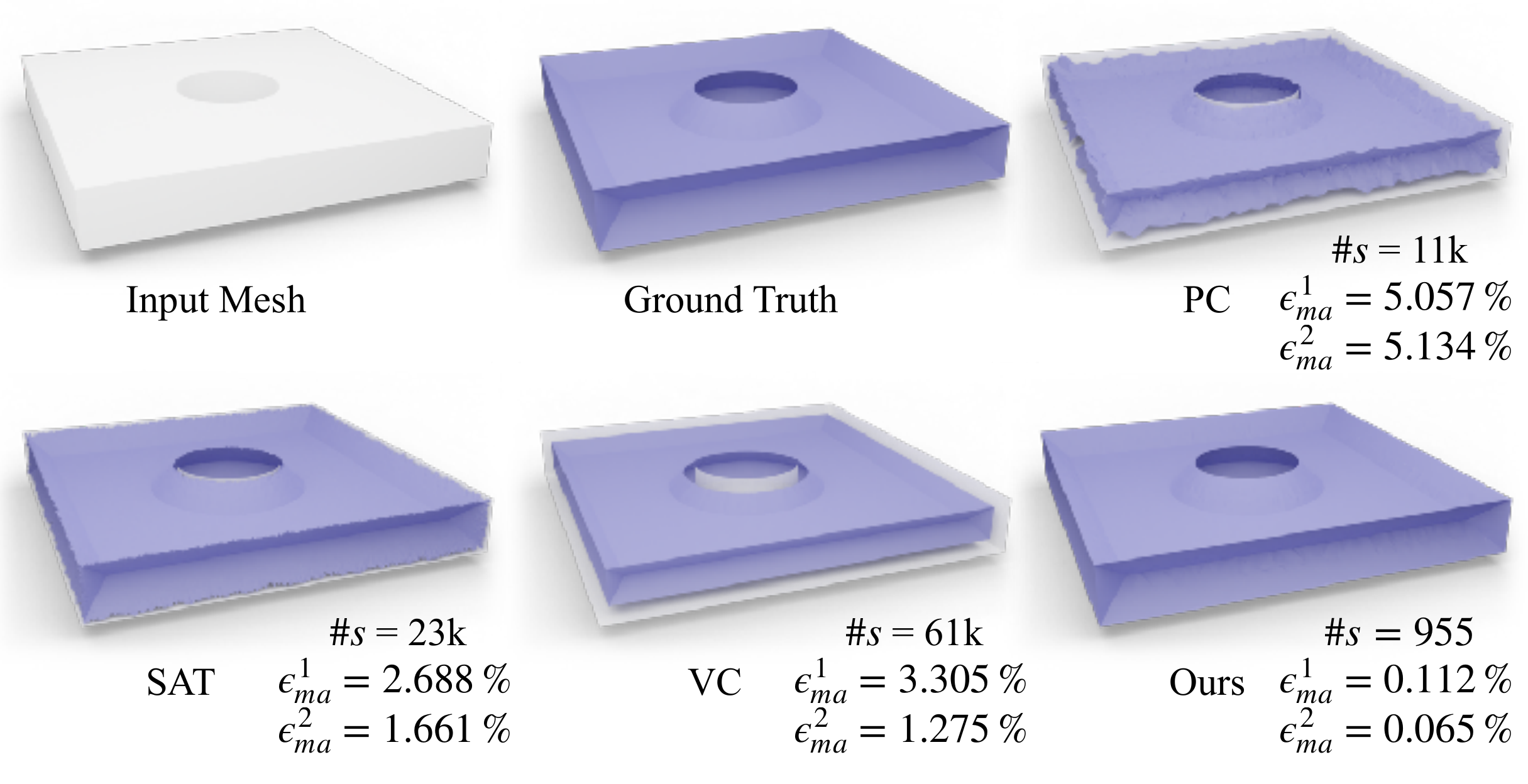}
    \vspace{-5pt}
    \caption{Quantitative comparison with PC \cite{amenta2001power}, SAT \cite{miklos2010discrete}, and VC \cite{yan2018voxel} on the medial mesh, with their Hausdorff distances measured w.r.t. the ground-truth medial axis.}
    \label{fig:comp_gt}
\end{figure}

We compare our method with Power Crust (PC) \cite{amenta2001power} regarding the generated medial mesh (Fig.~\ref{fig:comp_gt}), the surface reconstruction from MAT (Fig.~\ref{fig:comp_recon}), and the feature preservation quality (Fig.~\ref{fig:comp_features}). Similar to other sampling-based methods, the quality of medial mesh generated using PC would improve when the surface sampling density increases. However, their method cannot preserve any medial feature and the generated medial mesh is not thin with large number of flat tetrahedrons.

We also experiment with the SAT method \cite{miklos2010discrete} using two values of the sampling distance parameter: $\delta = 0.03$ and $\delta=0.01$, and setting the scale parameter to $s=1.0$. The smaller $\delta$ yields a good reconstruction precision. However, it favors a dense representation with a large number of medial spheres. The qualitative and quantitative comparison results are shown in Fig.~\ref{fig:comp_recon}.
SAT cannot preserve external features of input mesh surface in their MA results as the medial structure is not complete around convex sharp edges of input surfaces. Even though SAT generates promising result when preserving internal features on some models, it requires large amount of medial spheres (\ie 256k and 127k in Fig.~\ref{fig:comp_features} with $\delta=0.01$) comparing to our method (\ie 15k and 8k in Fig.~\ref{fig:comp_features}). When the number of medial spheres is not adequate even with same sampling parameter (\ie 7k and 57k in Fig.~\ref{fig:comp_recon} with $\delta=0.01$), SAT generates ill-posed faces around internal features. In addition, SAT also routinely produce topological errors, as shown by the Euler characteristic in Fig.~\ref{fig:comp_features}.

In contrast, our method preserves better external features and comparable (if not better) internal features, which leads to better surface reconstructions from the generated medial meshes. It is worth mentioning that, our method does not require large amount of medial spheres, since the design of our framework allows us to directly sample feature spheres without increasing the number of non-feature medial spheres as most of sampling-based and voxel-based methods do. In addition, the Euler characteristic shows the topological correctness and thinness of our generated medial mesh.

\subsection{Comparison with the \emph{Voxel Core} Method}

\begin{figure}[h]
    \centering
    \includegraphics[width=\linewidth]{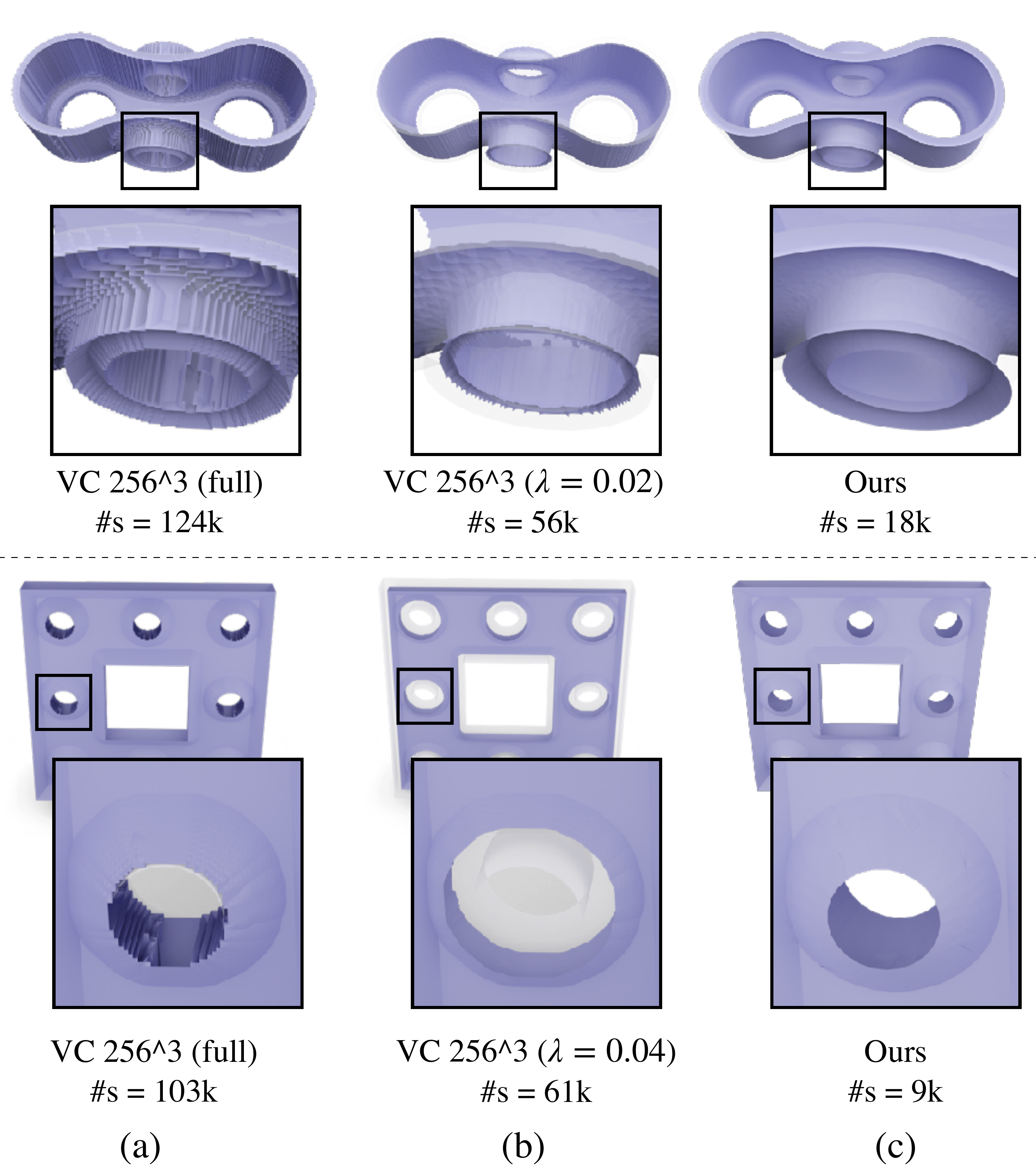}
    \vspace{-5pt}
    \caption{Comparing medial mesh results of our method (c) with VC \cite{yan2018voxel} w/o (a) and w/ (b) $\lambda$-pruning.}
    \label{fig:comp_vc_prune}
\end{figure}

We compare the reconstruction quality with the Voxel Core (VC) method \cite{yan2018voxel} by setting two voxel sizes: (1) $256^3$ with a default pruning parameter $\lambda=0.04$; (2) $512^3$ with pruning parameter $\lambda=0.01$ (the default parameter $0.04$ prunes excessively so that the medial mesh can not be properly maintained), as shown in Fig.~\ref{fig:comp_recon}. 
We also show a comparison of two VC results w/o and w/ pruning in Fig.~\ref{fig:comp_vc_prune}. 

The medial mesh generated from VC has the following two problems regarding external features. First, the more VC shrinks, the reconstructed shape is more rounded (Fig.~\ref{fig:comp_recon}). If VC shrinks less, redundant structures (Fig.~\ref{fig:comp_vc_prune} (a)) remain in the resulting medial mesh. Secondly, the boundary curves on VC's medial mesh have good correspondence with the external feature only when the sharp edges of input shape are parallel to the voxels’ boundaries. For external features of curves, VC produces zig-zag structures around the feature curves (Fig.~\ref{fig:comp_vc_prune} (b)). Similar as SAT, VC requires more number of medial spheres (usually $\geq 2$ times more) than ours in order to generate smooth medial structure around internal features (Fig.~\ref{fig:comp_vc_prune}). Our method, on the contrary, preserves the complete medial structure with a lower reconstruction error and a fewer number of medial spheres (Fig.~\ref{fig:comp_vc_prune} (c)).

\begin{figure}[h]
    \centering
    \includegraphics[width=0.95\linewidth]{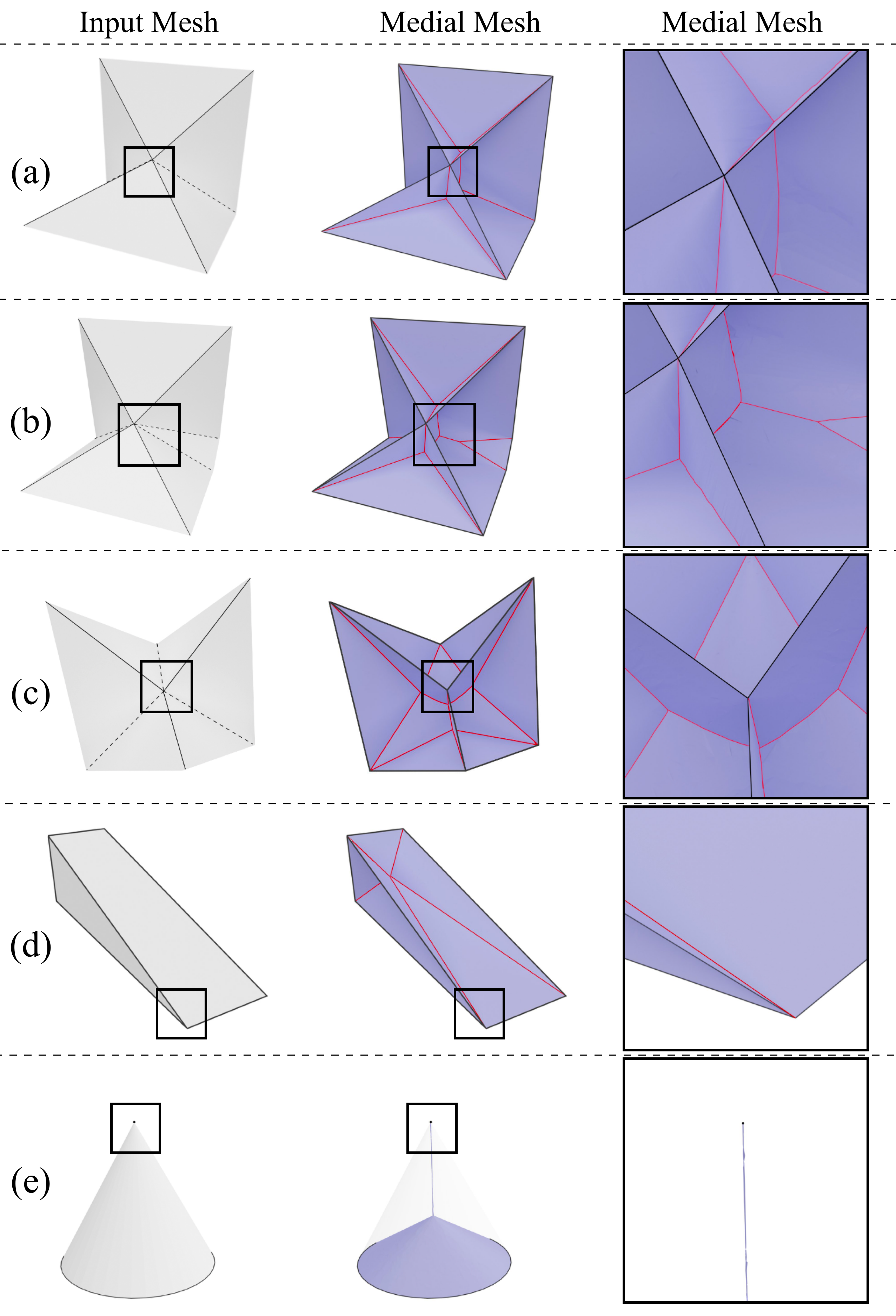}
    \vspace{-5pt}
    \caption{Examples of corner preservation in our medial mesh: (a) a corner adjacent to four convex edges and two concave edges, forming a saddle; (b) a corner adjacent to four convex edges and three concave edges (two of them are adjacent); (c) a corner adjacent to three convex edges and three concave edges, forming a monkey saddle; (d) a corner adjacent to three sharp edges at small angles, forming a wedge; and (e) a corner at the tip of a discretized cone.}
    \label{fig:saddle}
\end{figure}

\subsection{Ablation Study on Thinning Algorithm}
\label{sec:abl_study}

One important property of medial axis is its thinness, \ie it contains no three-dimensional cells. In this subsection, we give an in-depth analysis of the rational of the RPS-based sorting in our geometry-guided thinning algorithm.

\begin{figure}[h]
    \centering
    \includegraphics[width=\linewidth]{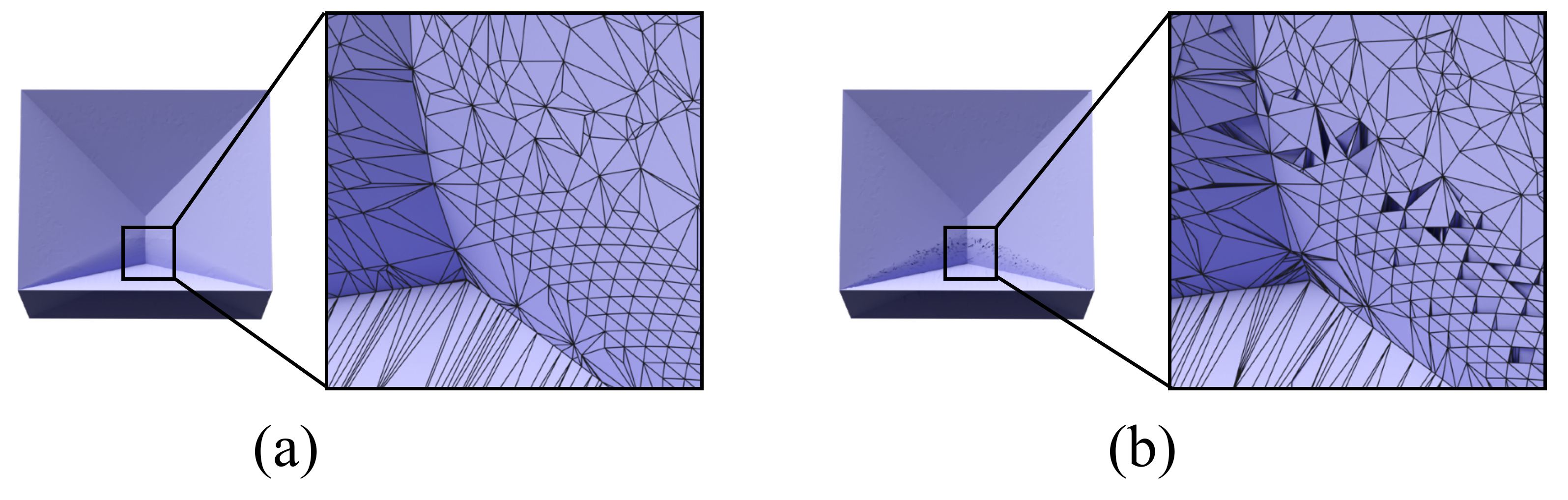}
    \vspace{-10pt}
    \caption{Comparison of two different thinning strategies on the model of Fig.~\ref{fig:saddle} (a): (a) medial faces sorted by our importance factor $\alpha_{ijw}$ (details in Sec.~\ref{sec:refine} and algorithm provided in Alg.~2 of Supplementary Material;
    (b) medial faces sorted randomly. We can see that plain thinning strategy (b) creates many open ``pockets'', while our geometry-guided thinning (a) produces an accurate medial mesh.}
    \label{fig:ablation_thinning}
    \vspace{-5pt}
\end{figure}

The plain thinning algorithm \cite{ju2007editing} prunes tetrahedrons by removing tet-face pairs randomly as long as they are simple pairs (\ie the face is on the boundary of only one tetrahedron). We found this operation routinely produce open ``pockets'' (triangles forming open cavities) even though the topology is correct. This is because the choice of simple pairs does not consider the geometry information as some faces are more important than others in a single tetrahedron. An example is shown in Fig.~\ref{fig:ablation_thinning}. The plain thinning algorithm creates unwanted open ``pockets'' (Fig.~\ref{fig:ablation_thinning} (b)) while our geometry-aware thinning algorithms produces geometrically-accurate medial meshes (Fig.~\ref{fig:ablation_thinning} (a)).

\begin{figure}[h]
\center
  \includegraphics[width=0.95\linewidth]{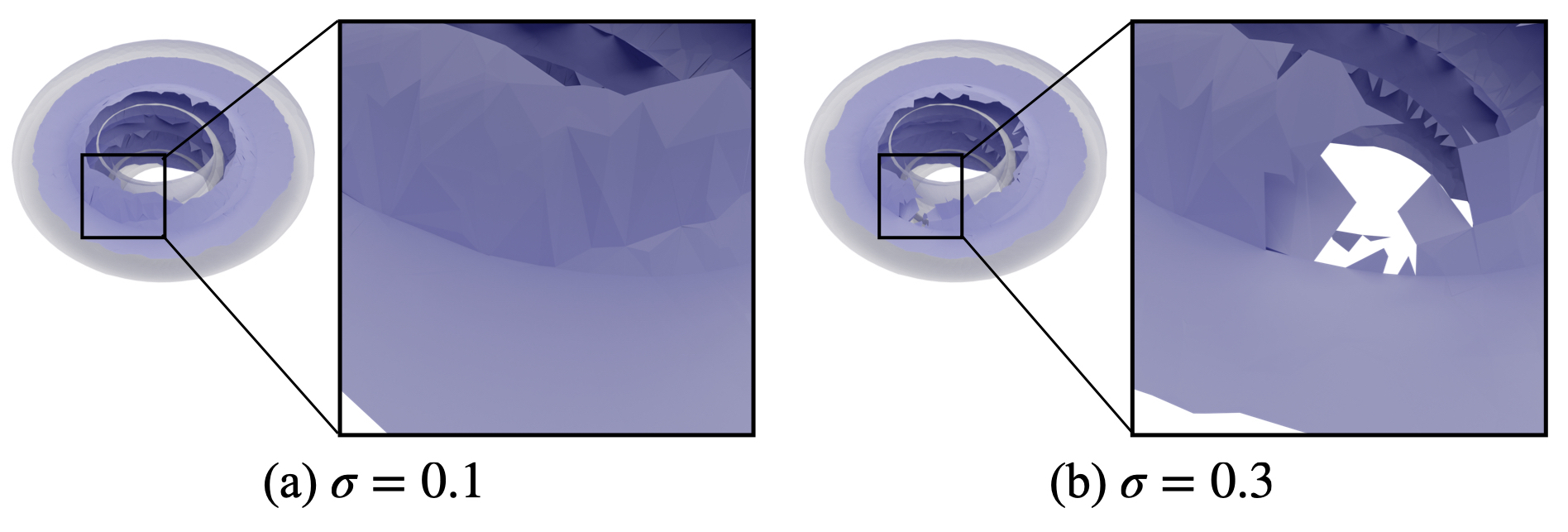}
  \caption{Thinning results using different target importance factor $\sigma$. Default value $\sigma=0.3$ (b) over-pruned the medial mesh which results in holes and detached components, while a smaller value of $\sigma=0.1$ (a) prevents the over-pruning.}
  \label{fig:thinning_param}
  \vspace{-10pt}
\end{figure}

The target importance factor $\sigma$ (in Sec.~\ref{sec:refine}) is a parameter that controls the simplicity of the output medial mesh. All tet-faces pairs will be removed no matter what value $\sigma$ is, and $\sigma$ plays as a stop sign for removing face-edge pairs. We show the effect of different values of the parameter $\sigma$ in Fig.~\ref{fig:thinning_param}. For medial mesh with sharp edges as closed boundaries, a larger value of $\sigma$ results in a cleaner medial mesh. We use $\sigma=0.3$ for all models in Table.~\ref{tab:more_rec_error}. For medial meshes whose boundaries are not closed sharp edges, however, the default value of $\sigma=0.3$ would remove face-edge pairs that are not suppose to be deleted. To prevent undesired deletion, we use $\sigma=0.1$ for models whose medial mesh does not ends at sharp features, which includes models in Fig.~\ref{fig:comp_non_feature} and models marked $\star$ in Tab.1 of Supplementary Material.

\begin{figure*}[h]
\center
  \includegraphics[width=\textwidth]{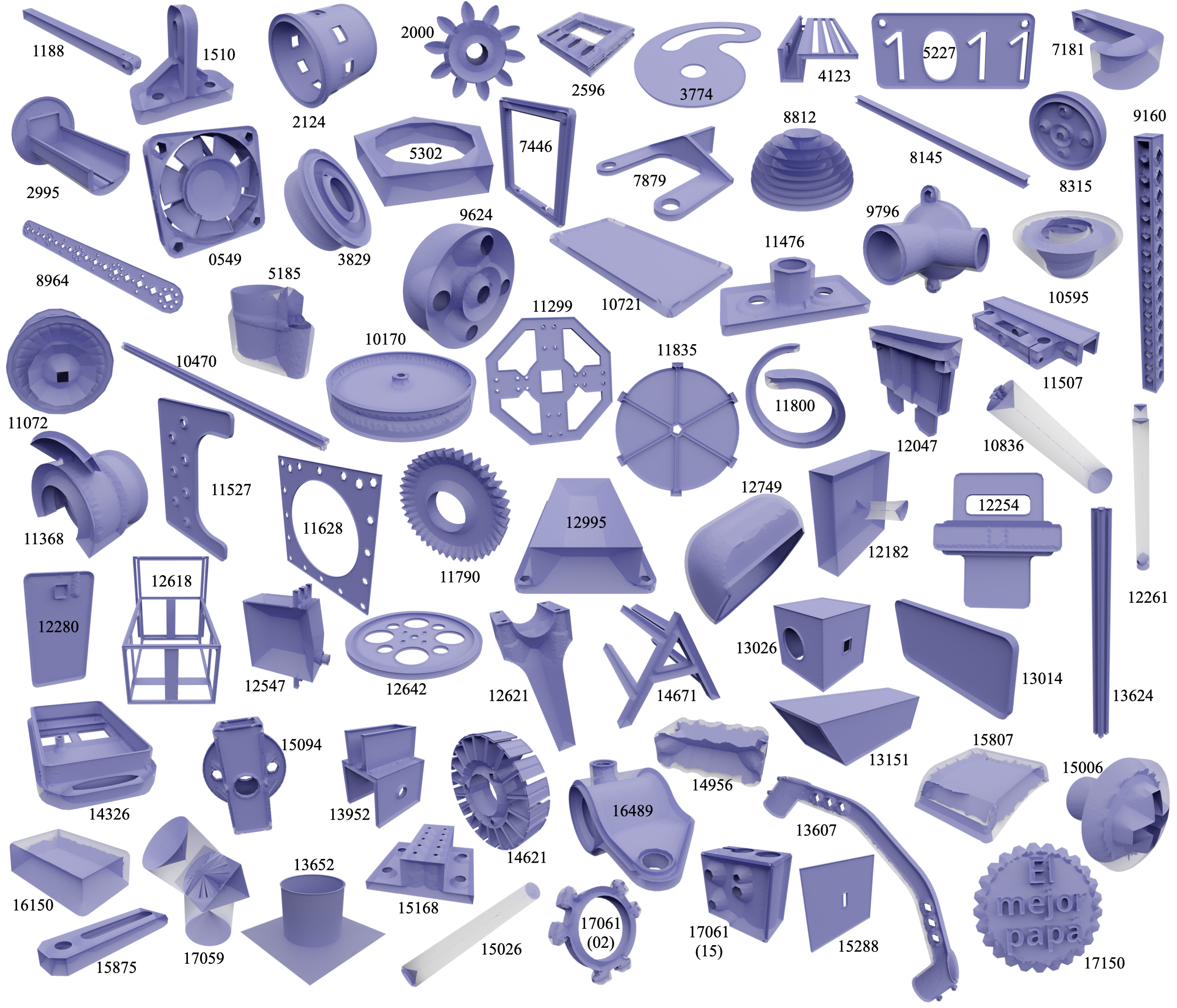}
  \vspace{-5pt}
  \caption{A gallery of our results using models from ABC dataset under the \textit{10k/test/2048} folder. The selection and statistics details are given in the Supplementary Material in Sec.~1 and Tab.~1.}
  \label{fig:gallery}
\end{figure*}

\begin{figure*}[h]
    \centering
    \includegraphics[width=\linewidth]{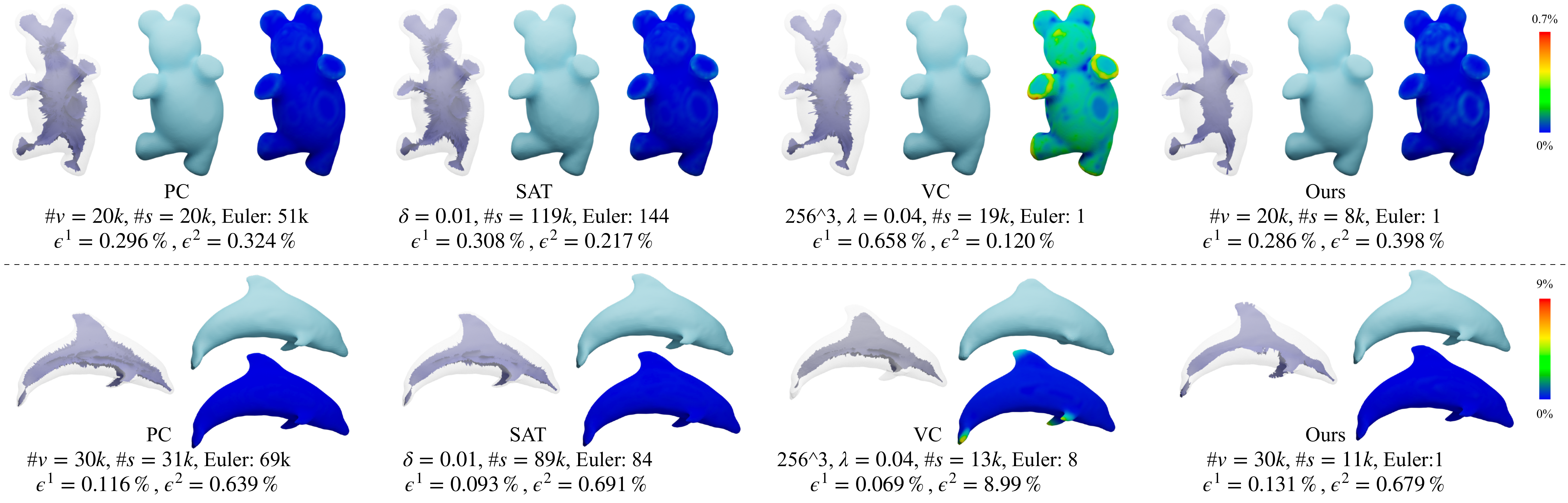}
    \vspace{-5pt}
    \caption{Comparison with PC \cite{amenta2001power}, SAT \cite{miklos2010discrete}, and VC \cite{yan2018voxel} on two non-CAD models. We show the generated medial meshes and the reconstructed surfaces, and the color-coded distribution of Hausdorff errors from the reconstructed surface to the input surface.}
    \label{fig:comp_non_feature}
\end{figure*}



\subsection{More Results}
\label{sec:results_more}

\paragraph{Results on Various Corner Features.}
We show five example results of our generated medial mesh of various corners in Fig.~\ref{fig:saddle}. The first three models (a), (b) and (c) contain saddles incident to both convex and concave edges. The fourth model (d) is a wedge with small angles. The fifth model (e) is a discretized cone with a corner at the tip. Since all of existing methods are known to have difficulties on preserving external corner features, we do not show their results as comparison. 

\paragraph{Results on Non-feature Shapes.}
We show a visual and quantitative comparison of our method with PC \cite{amenta2001power}, SAT \cite{miklos2010discrete} and VC \cite{yan2018voxel} on two smooth shapes in Fig.~\ref{fig:comp_non_feature}. We found that PC, SAT and our method normally give similar reconstruction accuracy while VC has higher errors since it shrank the generated medial mesh during pruning. Similar to models with features, our method generates fewer medial spheres (\ie PC  $20k$, SAT $119k$, and ours $8k$ for the bear model) and maintains the thinness property of medial axis while PC and SAT cannot.

We also show a gallery of more results from ABC dataset \cite{Koch_2019_CVPR} under the \textit{10k/test/2048} folder (see Fig.~\ref{fig:gallery}). Please refer to the Supplementary Material for detailed description regarding the selection (Sec.~1) and statistics (Tab.~1).

\section{Limitations and future work}
\label{sec:limitations}


It should be noted that our current approach does not guarantee topological preservation for the generated medial mesh w.r.t. the input model, as evidenced from Tab.~1 of Supplementary Material that there are still $19/73$ models having incorrect Euler characteristics. 
We need to investigate the necessary and sufficient conditions for topological equivalence under the RPD framework, and come up with some delicate mechanism to preserve the topology of the generated medial mesh. We will leave these topological investigations as our future work.

In addition to the topological preservation issue mentioned above, in this paper we only give experimental evidences but not theoretical proof of correctness for the proposed algorithms, which include: (1) the capability of RPC-based refinement in guaranteeing the topological correctness of internal features; (2) the topological correctness of corner feature preservation; (3) the geometric relationship between different choices of sphere connections (\ie medial triangles) and their RPS used in our geometry-guided thinning algorithm. 
More rigorous theoretical guarantees and limitations are needed to be explored in the future.



In addition, the preservation of external features can be effectively evaluated using reconstruction error, however, the same is not obvious for internal features. Moderate errors can be well hidden by the neighboring medial spheres when it comes to shape reconstruction. In our future work, we will consider evaluating the internal features on top of many hex-meshing applications which explicitly rely on internal features to perform solid-meshing of CAD models \cite{sampl2000semi} \cite{quadros2004laytracks}.

\section{Conclusion}

In this paper, we present a novel RPD-based framework for computing the medial axis transform of 3D shapes with preservation of both external and internal medial features. The method is based on the observation that the surface RPC of each medial sphere indicates the set of connected components (CCs) that the sphere has tangential contacts with. Each sphere's CCs can not only be used to update the spheres to their ground truth position and radius, but also tell the information about whether this sphere is on a medial sheet, a seam, or a junction. Such  information can be further used to check if the internal or external features are broken in the generated medial mesh, and guide the sphere sampling to preserve those features. Experimental evidences show that our method generates medial meshes with high quality in preserving medial features, both externally and internally. In the future, we believe our RPD-based framework, as a general tool for encoding shapes with features, has potential to be applied to various applications such as shape segmentation~\cite{lin2020seg}, shape recognition~\cite{Hu2019MATNet}, and shape deformation~\cite{Lan2020MedialElastics}, etc.

\begin{acks}
We would like to thank Shibo Song for helping us prepare figures and video for this paper. Ningna Wang and Xiaohu Guo were partially supported by National Science Foundation (OAC-2007661). Bin Wang was partially supported by National Key Research and Development Program of China (2020YFB1708900).
\end{acks}

\bibliographystyle{ACM-Reference-Format}
\bibliography{reference}


\end{document}